\documentclass[floatfix,superscriptaddress,a4paper,
               showpacs,showkeys,nofootinbib,preprint]{revtex4}
\usepackage{epsfig}
\usepackage{latexsym}
\usepackage{xspace}
\usepackage{hyperref}
\usepackage{inputenc}
\usepackage{indentfirst}
\usepackage{enumerate}
\usepackage{color}

\usepackage{amsmath}
\usepackage{amssymb}
\usepackage[english]{babel}
\usepackage{url}
\newcommand{\eq}[1]{\begin{align} #1 \end{align}}

\newcommand{\eV}{\ensuremath{\mbox{e\kern-0.1em V}}\xspace}
\newcommand{\GeV}{\ensuremath{\mbox{Ge\kern-0.1em V}}\xspace}
\newcommand{\MeV}{\ensuremath{\mbox{Me\kern-0.1em V}}\xspace}
\newcommand{\GeVc}{\ensuremath{\mbox{Ge\kern-0.1em V}\!/\!c}\xspace}
\newcommand{\GeVcc}{\ensuremath{\mbox{Ge\kern-0.1em V}\!/\!c^2}\xspace}
\newcommand{\AGeV}{\ensuremath{A\,\mbox{Ge\kern-0.1em V}}\xspace}
\newcommand{\AGeVc}{\ensuremath{A\,\mbox{Ge\kern-0.1em V}\!/\!c}\xspace}
\newcommand{\MeVc}{\ensuremath{\mbox{Me\kern-0.1em V}/c}\xspace}

\newcommand{\pip}{\ensuremath{\pi^+}\xspace}

\newcommand{\kp}{\ensuremath{\textup{K}^+}\xspace}

\newcommand{\NASixtyOne}{NA61\slash SHINE\xspace}

\newcommand{\CernVM}{\textsc{Cern\-\kern-0.05emVM}\xspace}

\begin{document}

\title{ Brief history of the search for critical structures \\ in heavy-ion collisions }

\author{Marek Gazdzicki}
\affiliation{Geothe-University Frankfurt am Main, Germany}
\affiliation{Jan Kochanowski University, Kielce, Poland}

\author{Mark Gorenstein}
\affiliation{Bogolyubov Institute for Theoretical Physics, Kiev Ukraine}

\author{Peter Seyboth}
\affiliation{Jan Kochanowski University, Kielce, Poland}

\begin{abstract}

The paper briefly presents history, status, and plans of the search for the critical structures - the onset of fireball, the onset of deconfinement, and the deconfinement critical point - in high energy nucleus-nucleus collisions. First, the basic ideas are introduced, the history of the observation of strongly interacting matter in heavy ion collisions is reviewed, and the path towards the quark-gluon plasma discovery is sketched. Then the status of the search for critical structures is discussed - the discovery of the onset of deconfinement, indications for the onset of fireball, and still inconclusive results concerning the deconfinement critical point. Finally, an attempt to formulate priorities for future measurements - charm quarks vs the onset of deconfinement and detailed study of the onset of fireball -- closes the paper.

\end{abstract}


\maketitle

\section{ Introduction and Vocabulary }
\label{sec:vocabulary}

One of the important issues of contemporary physics is the
understanding of strong interactions and in particular the study
of the properties of \textbf{strongly interacting matter} -- a system of strongly interacting particles in equilibrium.
The advent of the quark model of hadrons and the development of 
the commonly accepted theory of strong interactions, \textbf{quantum chromodynamics (QCD)},
naturally led to expectations that matter at very high densities
may exist in a state of quasi-free quarks and gluons, the \textbf{quark-gluon
plasma (QGP)}.

\begin{figure}[hbt]

\centering

 \includegraphics[width=0.99\linewidth,clip=true]{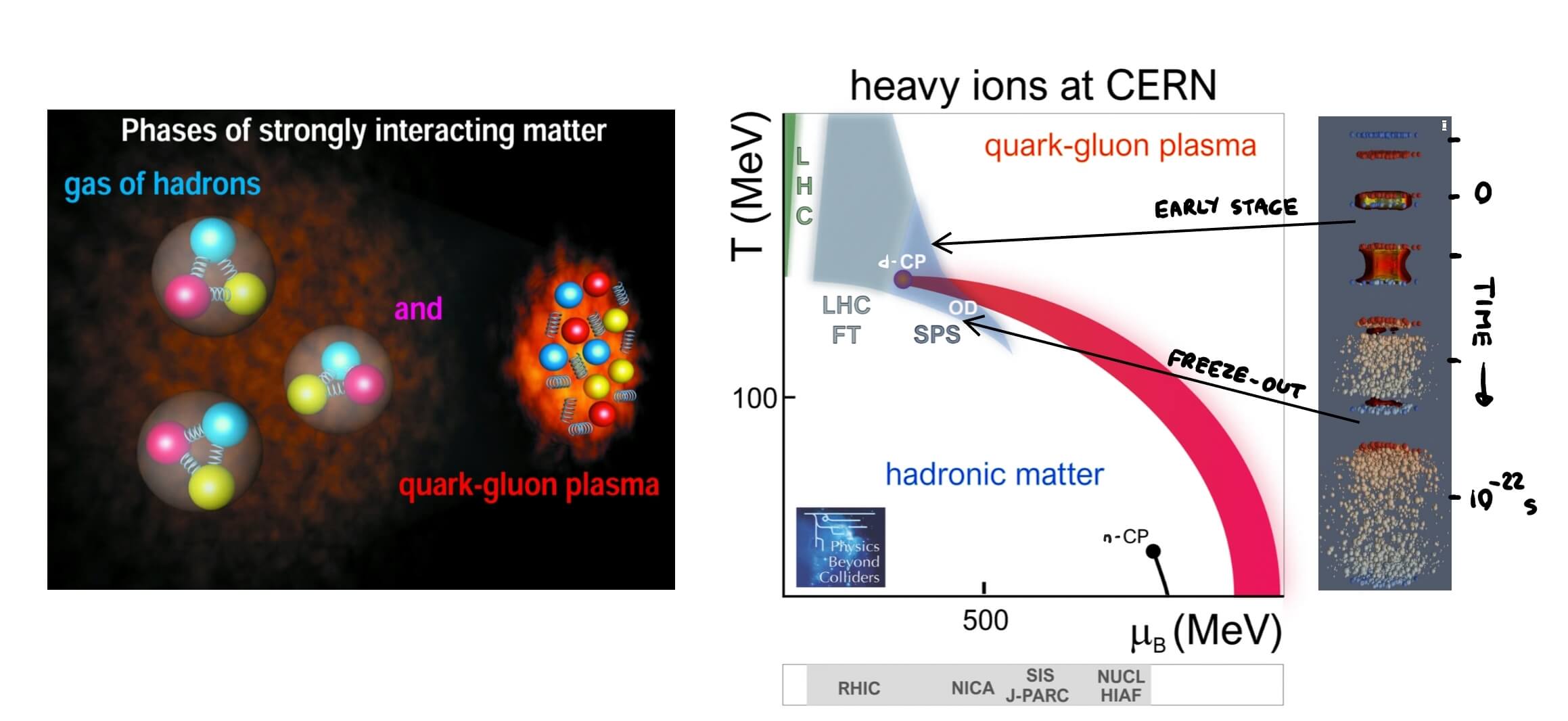}

\caption{
\textit{Left:} 
 Artistic sketch of the two phases of strongly interaction matter, hadron-resonance gas and quark-gluon plasma.
 \textit{Middle:}
Phase diagram of QCD in temperature $T$ and baryon chemical potential $\mu_B$, and the region covered by running or planned experiments~\cite{Alemany:2019vsk}. The density range covered by LHC, LHC-FT and SPS experiments is indicated by the shaded areas in the figure. The lower boundary of the grey and blue shaded area follows the chemical freeze-out.
The upper boundary relates to the parameters at the early stage of the collisions. The potential deconfinement critical point is labelled with d-CP, the onset of deconfinement with OD. The black line at small temperatures and high densities shows the nuclear liquid-gas transition, also ending in a critical point n-CP. The density range of other experiments is indicated in the bar below the figure. This
includes RHIC at BNL, NICA at JINR, SIS100 at FAIR, J-PARC-HI at J-PARC, the Nuclotron at JINR (NUCL), and HIAF at HIRFL.
 \textit{Right:} 
  Evolution of a heavy-ion collision at high energies. Successive snapshots of a central collision are shown versus time.  
 }
 \label{fig:v_qgp}
\end{figure}

Does the QGP exist in nature?
How does the transition proceed from a low-density state of strongly interacting matter, 
in which quarks and gluons are confined in hadrons, to the QGP? 
Is it similar to the transition from liquid water to water vapour along
a first order transition line ending in a second order critical point and followed by a
cross over transition, see illustration plots in Fig.~\ref{fig:v_qgp}?

The study of high energy collisions of two atomic nuclei gives us
the unique possibility to address these issues in well controlled
laboratory experiments. 
This is because it is observed that a system of strongly 
interacting particles created in \textbf{central heavy ion collisions} is close to (at least local) equilibrium.
How does the transition from a non-equilibrium system created in \textbf{inelastic proton-proton interactions}
to the equilibrium system in central heavy ion collisions look like?

These questions have motivated broad
experimental and theoretical efforts for about 50 years. 
Systematic measurements of particle production properties in nucleus-nucleus (A+A) collisions at different collision energies and for different masses of
colliding nuclei have been performed.
By changing collision energy and nuclear mass number 
one changes macroscopic parameters of the created system -- its volume, energy, and net baryon number.
This allows to move across the phase diagram and look for the theoretically predicted boundaries of equilibration and matter phases, see  illustration plots in Fig.~\ref{fig:v_qgp}.
Consequently several physics phenomena might be 
observed when studying experimentally nuclear collisions at high energies. 
These are:
\begin{enumerate}[(i)]
\item 
\textbf{onset of fireball} -- beginning of creation of large-volume ($\gg 1$~fm$^3$) strongly interacting matter,

\item
\textbf{onset of deconfinement} -- beginning of QGP creation with increasing collision energy,

\item 
\textbf{deconfinement critical point}  -- a hypothetical end point of the first order transition line to quark-gluon plasma that has properties of a second order phase transition. 
\end{enumerate} 

These phenomena are expected to lead to rapid changes of hadron production properties -- the \textbf{critical structures} -- 
when changing collision energy and/or nuclear mass number of the colliding nuclei.

\section{ Strongly interacting matter in heavy ion collisions}
\label{sec:strongly}


\textbf{Strongly interacting matter.}
The equation of state defines the macroscopic properties of matter in equilibrium. It  is a subject of statistical mechanics. The first step in this modelling is to clarify  the types of particle species and 
inter-particle interactions.
One should also choose an appropriate statistical ensemble
which fixes the boundary conditions and conserves the corresponding global physical quantities, like energy and conserved charges.
Strongly interacting matter at high energy density can be formed at the early stages of
relativistic 
A+A collisions.
As mentioned in the introduction the questions --
\begin{enumerate}[(i)]
\item
What types of particles should be considered as fundamental?
\item
What are the composite objects?
\item
What are the fundamental forces between the matter constituents?
\item
What are the conserved charges?
\end{enumerate}
-- should be addressed.
%
Answers to these questions are changing with time as our knowledge about basic  properties of elementary particles and their interactions increases.

The first model of strongly interacting matter at high energy density was formulated in 1950 by Fermi~\cite{fermi:1950jd}.
It assumes that a system created in high energy proton-proton (p+p) interactions emits  pions like black-body radiation, i.e., pions are treated as non-interacting particles,
and the pion mass $m_\pi\cong 140$~\MeV is neglected compared to the high temperature of the system.
The pressure $p$ and energy density $\varepsilon$ can then be represented by the following functions of the temperature $T$ 
(the system of units with $h/(2\pi)=c=k_B=1$ will be used),
\eq{\label{p}
p(T)~=~\frac{\sigma}{3}T^4~,~~~~~\varepsilon(T)~\equiv~T\frac{dp}{dT}~-~p~=\sigma T^4~,
}
where $\sigma= \pi^2g/30$ is the the so-called Stephan-Boltzmann constant, with $g$ being the degeneracy factor (the number of spin and isospin states), and $g=3$ counting the three isospin states $\pi^+,~\pi^0,~\pi^-$.  

\vspace{0.2cm}
\textbf{Hadrons and resonances.}
The study of particle production in high energy collisions started in the 1950s with discoveries
of the lightest hadrons -- $\pi$, $K$, and $\Lambda$
-- in cosmic-ray experiments. Soon after with the rapid advent of particle accelerators
new particles were discovered almost day--by--day.
The main feature of the strong interactions appears to be creation of new and new types of particle species -- hadrons and resonances -- when increasing the collision energy. 
A huge number (several hundreds) of different hadron and resonance species are known today. The simplest statistical model treats the hadron matter, i.e. a system of strongly interacting particles at not too high energy density, as a mixture of ideal gases of different hadron-resonance species.  

\vspace{0.2cm}
\textbf{Hadron-resonance gas.}
In the grand canonical ensemble the pressure 
function is then written as
%
\begin{eqnarray}
\label{p-id}
p^{\rm id}(T,\mu) 
= \sum_i\frac{g_i}{6\pi^2}\int d m\, f_i(m)\int_0^{\infty} \frac{k^4dk}{\sqrt{k^2+m^2}}
\left[ \exp\left(\frac{\sqrt{k^2+m^2} - \mu_i}{T}\right)+\eta_i\right]^{-1}~,
\end{eqnarray}
where $g_i$ is the degeneracy factor of the $i^{\textrm{th}}$ particle and the normalized
function $f_i(m)$
takes into account the Breit-Wigner shape of resonances with finite width
$\Gamma_i$ around their average mass $m_i$. For the stable hadrons,
$f_i(m)=\delta(m-m_i)$.
The sum over $i$ in Eq.~\eqref{p-id} 
is taken over all
non-strange and strange hadrons listed in the Particle Data Tables.
Note, that in the equation
$\eta_i =-1$ and $\eta_i = 1$ for bosons
and fermions, respectively, while $\eta = 0$ corresponds to the Boltzmann approximation.
The chemical potential for the $i^{\textrm{th}}$  hadron is given by
\begin{equation}
\mu_i\ =\ b_i\,\mu_B\, +\, s_i\,\mu_S\, +\, q_i\,\mu_Q
\label{eq:mui}
\end{equation}
with $b_i = 0,\, \pm 1$, $s_i = 0,\, \pm 1,\, \pm 2,\, \pm 3$, and
$q_i = 0,\, \pm 1,\, \pm 2$
being the corresponding baryonic number, strangeness, and electric charge of
the $i$th  hadron. Hadrons composed of charmed and beauty quarks are rather heavy and thus rare in the hadron-resonance gas, and their contribution to the thermodynamical functions are often neglected. 
Chemical potentials are denoted as $\mu\equiv (\mu_B,\mu_S,\mu_Q)$ and correspond to the conservation of net-baryon number, strangeness, and electric charge in the hadron-resonance gas.
The entropy density $s$,  net-charge densities $n_i$ (with $i=B,Q,S$), and energy density $\varepsilon$  are calculated from the pressure function (\ref{p-id}) according to the standard thermodynamic identities:
\eq{\label{therm-id}
s(T,\mu)=\left(\frac{\partial p}{\partial T}\right)_\mu~,~~~
n_i(T,\mu)=\left(\frac{\partial p}{\partial \mu_i}\right)_T~,~~~~ \varepsilon(T,\mu)=Ts+\sum_in_i\mu_i-p~.
}

Note that only one chemical potential $\mu_B$ is considered as independent variable in fits of the model to particle multiplicities produced in A+A reactions. Two others, $\mu_S$ and $\mu_Q$, should be found, at each pair of $T$ and $\mu_B$, from the requirements that the net-strangeness density equals to zero, $n_S=0$, and the ratio of the net-electric charge density, $n_Q$, to $n_B$ equals to the ratio of the number of protons, $Z$, to the number of all nucleons, $A$ (protons and neutrons) in the colliding nuclei, $n_Q/n_B=Z/A$.  
Equations (\ref{p-id}-\ref{therm-id}) define the ideal hadron-resonance gas  
model. In spite of evident simplifications this model  rather successfully fits the rich data on mean multiplicities of hadrons measured in central A+A collisions at high energies.  

\vspace{0.2cm}
\textbf{Hagedorn model.} 
Is there an upper limit for the masses of mesonic and baryonic resonances? 
In 1965 Hagedorn  formulated a statistical model assuming an exponentially increasing spectrum  of hadron-resonance states at large masses~\cite{Hagedorn:1965st}:
\eq{\label{H}
\rho(m)~\cong~C\,m^{-a}\,\exp\left(\frac{m}{T_H}\right)~, 
}
where $C$, $a$, and $T_H$ are the model parameters. 
At that time the number of experimentally detected hadron-resonance states was much smaller than it is today. Nevertheless, Hagedorn made the brave assumption that these $m$-states interpolate the low-mass spectrum and extend to $m\rightarrow \infty$, and that their density at large $m$ behaves as in Eq.~(\ref{H}).   
The pressure function at $\mu=0$ then becomes
\eq{ \label{pH}
p(T)~=~T\int_0^\infty dm\,\rho(m)\,\phi_m(T)~,
}
with the function $\phi_m(T)$ behaving at $m/T \gg 1$ as
\eq{ \label{large-m}
 \phi_m(T)~\cong ~ g\,\left(\frac{mT}{2\pi}\right)^{3/2}\,\exp\left(-\,\frac{m}{T}\right)~.
}
The result (\ref{large-m}) can be found from Eq.~(\ref{p-id}) after $k$-integration. At $m/T\gg1$ and $\mu=0$
both quantum statistics and relativistic effects become negligible.

 There are two exponential functions in the integrand (\ref{pH}):  $\exp(-m/T)$ defines the exponentially decreasing  contribution of each individual $m$-state, and $\exp(m/T_H)$ defines the exponentially increasing number of these $m$-states. It is clear that the $m$-integral
in Eq.~(\ref{pH}) exists only for $T \le T_H$.
Therefore, a new hypothetical
physical constant -- the limiting temperature
$T_H$ -- was introduced. The numerical value of $T_H$ was estimated by Hagedorn from two sources: from the straightforward comparison of Eq.~(\ref{H}) with the experimental mass spectrum of hadrons and resonances, $\Delta N/\Delta m$, and  from the inverse slope parameter of the transverse momentum spectra of final state hadrons in p+p interactions at high energies. Both estimates gave similar values $T_H=150-160$~MeV.  
The hadron states with large $m$ in the Hagedorn model are named the Hagedorn fireballs. These states were defined in a democratic (bootstrap) way:
the Hagedorn fireball consists of an arbitrary number of non-interacting Hagedorn fireballs, each of which in turn consists of ...

In the 1960s it was not clear up to what masses the hadron-resonances spectrum can be extended.  
The answer to this question is still unclear today. The large (exponential) density of resonance states
$\rho(m)$ and the finite widths $\Gamma(m)$ of these states  make their experimental observation very problematic.  
Moreover, several conceptual problems of the Hagedorn  model were obvious from the very beginning. The lightest hadron species, e.g., pion, kaon and proton can not be considered as  composed of other (non-interacting) hadrons, and should therefore have their own (non-democratic) status. Besides, the fireballs are treated as point like non-interacting  objects. However, from nuclear physics it was already evident that at least protons and neutrons are (strongly) interacting particles: nucleons should have both attractive and repulsive interactions to be able to form stable nuclei.  Most probably, similar interactions exist between other types of baryons. Evident physical arguments suggest that the same type of repulsive and attractive interactions should exist between anti-baryon species.

\vspace{0.2cm}
\textbf{Quark-gluon plasma.}
The quark model of hadron classification was proposed by 
Gell--Mann~\cite{GellMann:1964nj} and  Zweig~\cite{Zweig:352337} in 1964. It was the alternative to the bootstrap approach. Only three types of objects -- $u$, $d$, $s$ quarks and their anti-quarks -- were needed to construct the quantum numbers of  all known hadrons and successfully predict several new ones.  A 15 years 
period then started in which the idea of the existence of sub-hadronic particles -- quarks and gluons -- 
was transformed into the fundamental  theory of strong interactions,
quantum chromodynamics (QCD). Soon after the discovery of the $J/\psi$-meson in 1974 three new types of quarks -- $c$, $b$, and $t$ -- were added to QCD.  
In parallel, an important conjecture was formulated~\cite{Ivanenko:1965dg,Itoh:1970uw} --
matter at high energy density, as in super-dense star cores, may consist of quasi-free Gell-Mann-Zweig quarks instead of
densely packed hadrons. Some years later Shuryak investigated the properties of QCD matter and came to a qualitatively similar conclusion:  QCD matter at high temperature is best described by quark and gluon degrees of
freedom and the name quark-gluon plasma was coined~\cite{Shuryak:1977ut,Shuryak:1980tp}.

Questions concerning QGP properties and properties of its transition
to matter consisting of hadrons, have been considered since the late 1970s (see, e.g., Ref.~\cite{Cabibbo:1975ig}). The Hagedorn model was still rather popular at that time  due to its successful phenomenological applications. For example, the temperature parameter $T_{\rm ch}$ found from fitting the data on hadron multiplicities in p+p interactions and A+A collisions at high energies (the so-called chemical freeze-out temperature) was found to be close to the limiting Hagedorn temperature, $T_{\rm ch} = 140-160~\MeV \cong T_H $.  
Hagedorn and Rafelski~\cite{Hagedorn:1980cv}
as well as Gorenstein, Petrov, and Zinovjev~\cite{Gorenstein:1981fa} 
suggested that the 
upper limit of the hadron temperature, the Hagedorn
temperature $T_H$, is not the limiting temperature but the transition temperature
to the QGP, $T_C = T_H \approx 150$~\MeV.  
The Hagedorn fireball was then interpreted
as the quark-gluon bag formed in the early stage of the collision. It also had an exponential mass spectrum (\ref{H}) like in the Hagedorn model,
but was not a point like object. The average volume of the quark-gluon bag increases linearly with its mass.
This causes the excluded volume effects in the system of bags and leads to the transition to the high temperature QGP phase.
Note that the first QCD-inspired estimate of the transition temperature to the
QGP gave $T_C \approx 500~\MeV$~\cite{Shuryak:1977ut}, the most recent QCD-based estimates obtain $T_C \approx T_H \approx 150~\MeV$~\cite{Bazavov:2018mes}. 

Many physicists started to speculate that the QGP could be formed in
A+A collisions at sufficiently high energies in which one expects that strongly interacting matter of high energy density will be created.
Therefore the QGP might be discovered in laboratory experiments.


\begin{figure}[hbt]

\centering
 \includegraphics[width=0.99\linewidth,clip=true]{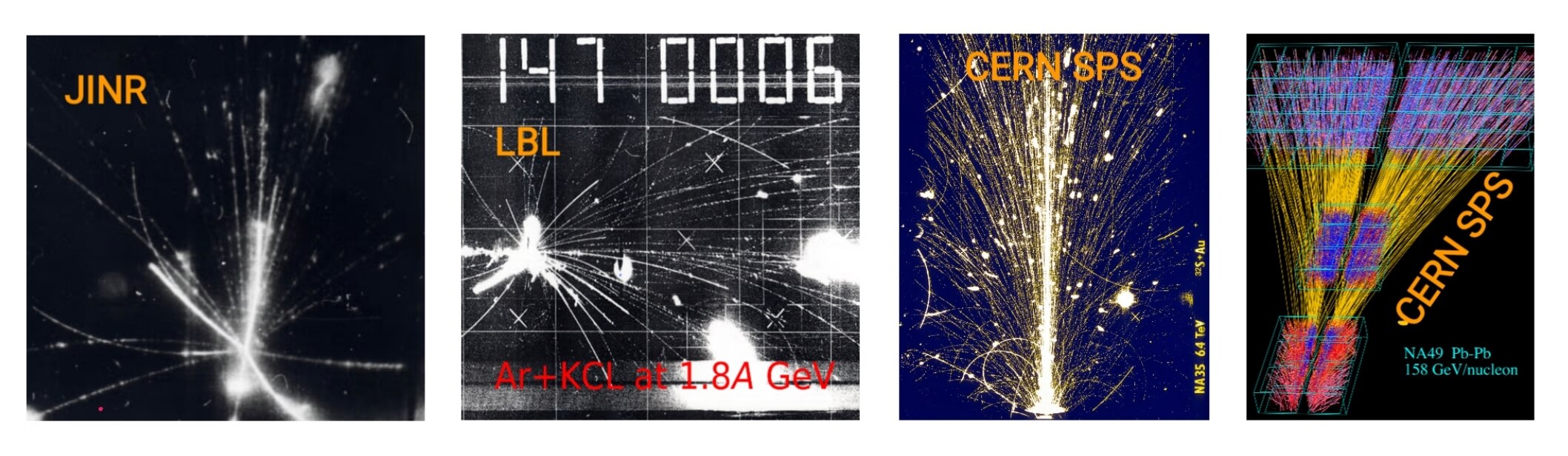}
\caption{
Tracks produced in nucleus-nucleus collisions recorded by heavy-ion experiments
located in JINR Dubna (the SKM-200 streamer chamber~\cite{Aksinenko:1980nm}), LBL Berkeley (the LBL streamer chamber~\cite{Sandoval:1982cn}), CERN SPS (the NA35 streamer chamber~\cite{Stock:1987rn}) and CERN SPS (the NA49 time projection chambers~\cite{Afanasev:1999iu}), from left to right, respectively. 
 }
\label{fig:s_events}
\end{figure}

\vspace{0.2cm}
\textbf{The first experiments.}
In parallel to the theoretical ideas and models, experimental studies of A+A
collisions were initiated in 1970 at the Synchrophasotron (JINR Dubna)~\cite{Issinsky:1994it,Malakhov:2013zjq} and in 1975 at the Bevelac (LBL Berkeley)~\cite{Alonso:1985awy}. Figure~\ref{fig:s_events} (\textit{first} and \textit{second from left}) shows two examples of recorded collisions.

Several effects were observed which could be attributed to a collective behaviour of the created system of hadrons. These are anisotropic and radial flow of particles~\cite{Lisa:1994yr,Nagamiya:1981sd,Gustafsson:1984nd}, enhanced production of strange particles~\cite{Anikina:1984cu} and suppressed production of pions~\cite{Sandoval:1980bm}. They could only be 
explained by
assuming that  strongly interacting matter is produced in the studied collisions~\cite{Stock:1985xe,Danielewicz:1985hn}. 
In what follows we use the term fireball as the notation for a large volume
($\gg 1$~fm$^3$) system consisting of strongly interacting particles close to at least local equilibrium. They can be either hadrons and resonances or quarks and gluons.

\vspace{0.2cm}
\textbf{Initiating the hunt for QGP.}
The two findings,
\begin{enumerate}[(i)]
\item
theoretical: QCD matter at sufficiently high temperature is in the state of a QGP;
\item
experimental: strongly interacting matter is produced in heavy ion collisions at energies of several GeV per nucleon;
\end{enumerate}
led activists of the field~\cite{Satz:2016xba} to the important decision to collide heavy ions at the maximum possible energy with the aim to discover the QGP.
In the 1980s the maximum possible energy for heavy ion collisions was available at CERN, Geneva.
This is why heavy ion physics entered the Super Proton Synchrotron (SPS) program at CERN.
The \textit{Workshop on future relativistic heavy ion experiments}, GSI Darmstadt, October 7-10, 1980, organized by Bock and Stock~\cite{Bock:1981iyr}, with an opening talk by Willis and a summary talk by Specht led to the formulation of the new program. Moreover, it initiated a series of \textit{Quark Matter} conferences~\cite{Satz:2016xba}.

\section{Evidence for the quark-gluon plasma}
\label{sec:q}

\textbf{Predicted QGP signals.}
The experimental search for a quark-gluon plasma in heavy ion collisions at the CERN SPS was shaped by several model predictions of possible QGP signals:
\begin{enumerate}[(i)]
  \item
    suppressed production of charmonium states, in particular $J/\psi$ mesons~\cite{Matsui:1986dk},
    \item
    enhanced production of strange and multi-strange hadrons from the QGP~\cite{Rafelski:1982pu},
     \item 
    characteristic radiation of photons and dilepton pairs from the QGP~\cite{Shuryak:1980tp}.
\end{enumerate}

\vspace{0.2cm}
\textbf{Measurements at the CERN SPS.}
The search for the QGP at the CERN SPS was performed in two steps:
\begin{enumerate}[(i)]
    \item 
    In 1986-1987 oxygen and sulphur nuclei were accelerated to 200\AGeV.
    Data on collisions with various nuclear targets were recorded by seven experiments, NA34-2, NA35, NA36, NA38, WA80, WA85 and WA94. 
    \item
    In 1996-2003 lead and indium beams at 158\AGeV were collided with lead and indium targets. Data were recorded by nine experiments, NA44, NA45, NA49, NA50, NA52, NA57, NA60, WA97 and WA98. 
\end{enumerate}

Figure~\ref{fig:s_events} shows a S+Au collision at 200\AGeV (\textit{second from right}) and a Pb+Pb collision at 158\AGeV (\textit{right}) recorded by the NA35 streamer chamber~\cite{Stock:1987rn} and the NA49 time projection chambers~\cite{Afanasev:1999iu}, respectively. 

\begin{figure}
    \centering
    \includegraphics[width=0.5\linewidth,clip=true]{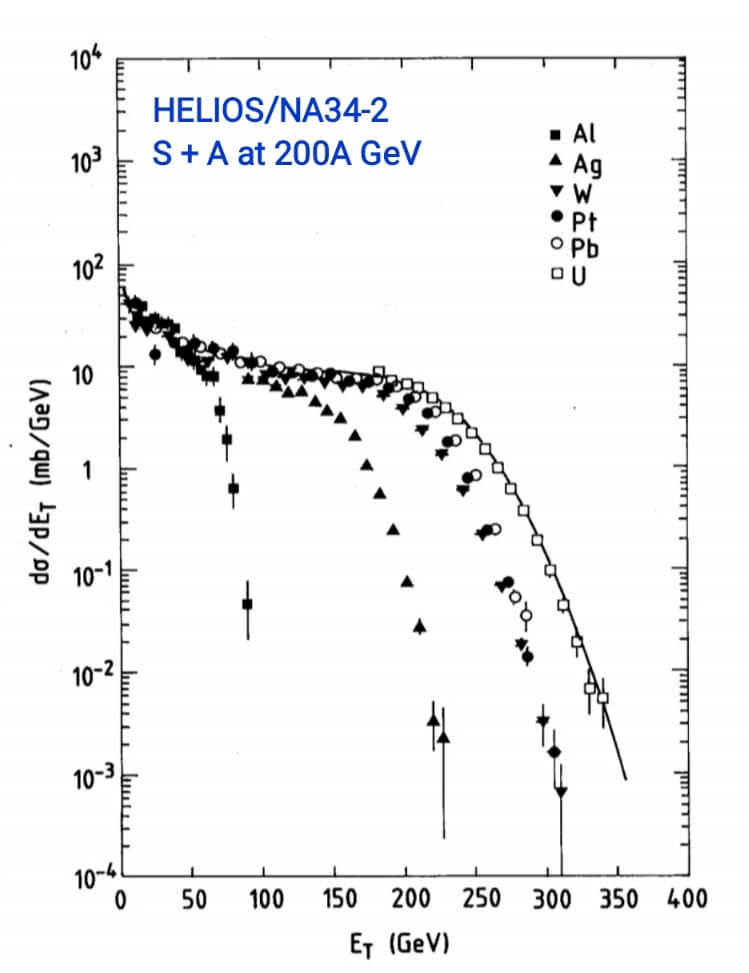}
        \caption{Transverse energy distributions in S+A collisions at the top CERN SPS energy measured by HELIOS/NA34-2~\cite{Akesson:1988yt}.}
    \label{fig:detadet}
\end{figure}

\begin{figure}
   \centering
    \includegraphics[width=0.48\linewidth,clip=true]{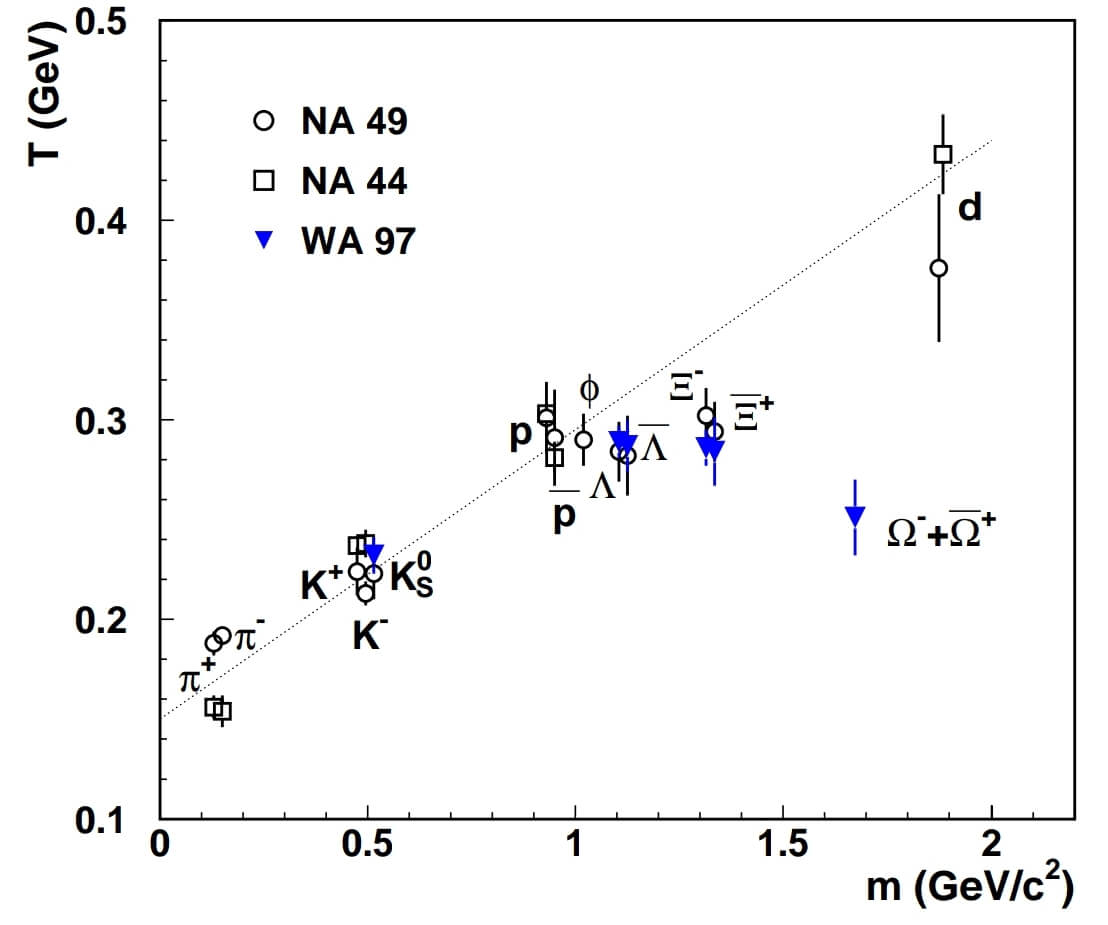}
    \includegraphics[width=0.51\linewidth,clip=true]{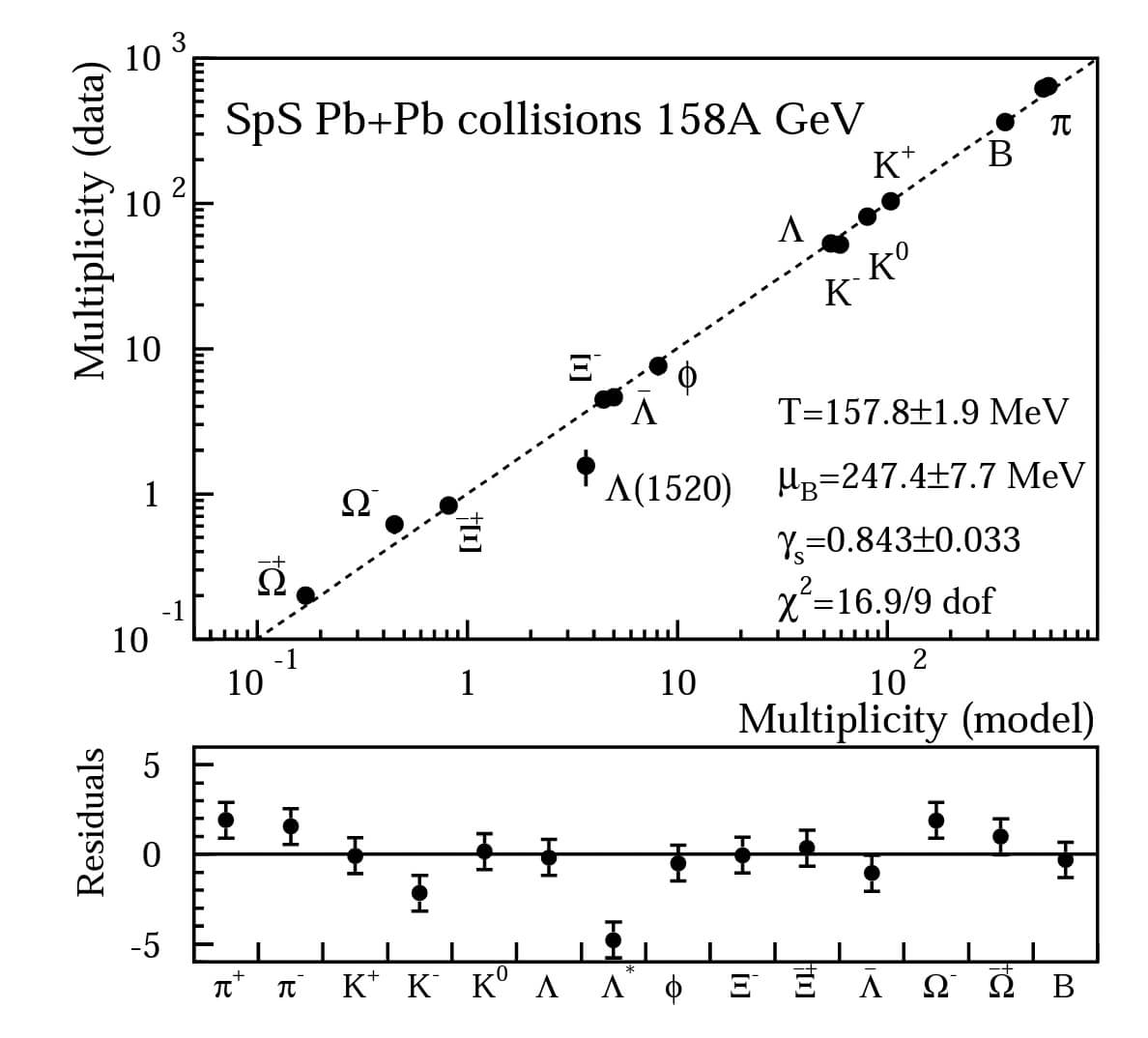}
    \caption{\textit{Left:} Inverse slope parameter (effective temperature) of the transverse mass distribution versus particle mass measured by WA97, NA44 and NA49~\cite{Antinori:2000sb}. \textit{Right:} Mean hadron multiplicities
    measured by NA49 compared to the statistical model fit~\cite{Becattini:2003wp}. Pb+Pb collisions at the top CERN SPS energy. }
   \label{fig:exp-stat}
\end{figure}

An estimate of the energy density in A+A
collisions can be obtained from measurement
of the transverse energy production and the size of the collision system. Already in 
S collisions with heavy nuclei (see Fig.~\ref{fig:detadet}) it was found that values above 1~GeV/fm$^3$ were
reached (NA34~\cite{Akesson:1987kh,Akesson:1988yt}, NA35~\cite{Heck:1988cm}, NA49~\cite{Margetis:1994tt}).
Moreover the fireball showed effective temperature increasing linearly with particle mass,
a characteristic of collective radial expansion (see Fig.~\ref{fig:exp-stat} (\textit{left})). Also mean
multiplicities of produced hadrons are well reproduced by the statistical model~\cite{Becattini:2003wp} (see Fig.~\ref{fig:exp-stat} (\textit{right})).
Thus conditions in collisions of heavy nuclei at the top energy of the CERN SPS are promising for the production of the QGP.

\begin{figure}[hbt]
\centering
 \includegraphics[width=0.99\linewidth,clip=true]{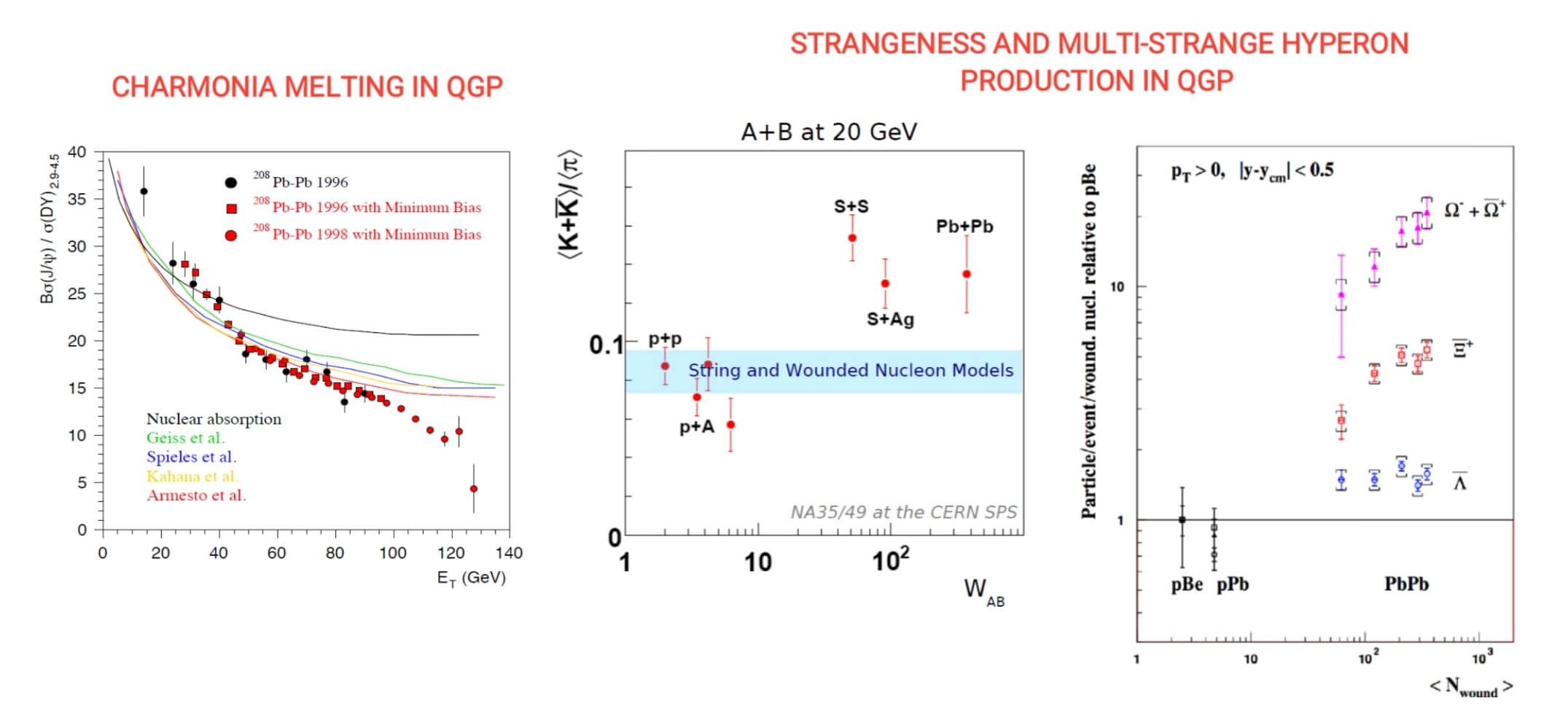}
\caption{
\textit{Left:} Ratio of $J/\psi$ meson to Drell-Yan muon pair production (data points) yields
compared to predictions (curves) of $J/\psi$ absorption by hadronic 
matter~\cite{Baglin:1990iv,Alessandro:2004ap} (NA38, NA50).
\textit{Center:} Comparison of \kp/\pip yield ratio in p+p, p+A and A+A collisions~\cite{Bartke:1990cn,Appelshauser:1998vn}
(NA35, NA49). \textit{Right:} Comparison of the mid-rapidity ratios of
of hyperon production to number of wounded nucleons in p+Be, p+Pb and Pb+Pb
collisions~\cite{Antinori:2006ij} (NA57). Top CERN SPS energy.
 }
\label{fig:q_signalsa}
\end{figure}

\begin{figure}[hbt]
\centering
 \includegraphics[width=0.4\linewidth,clip=true]{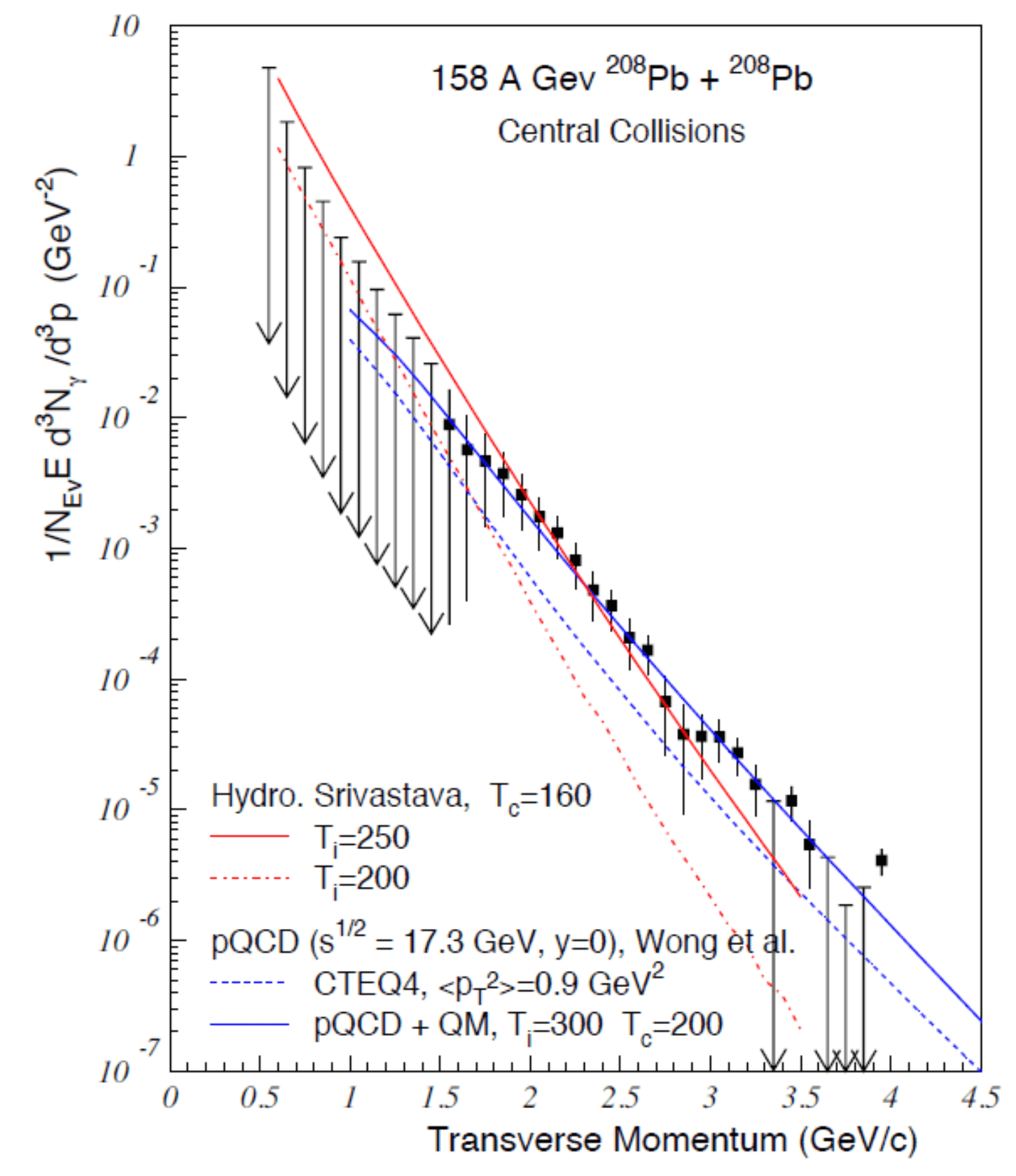}
 \includegraphics[width=0.4\linewidth,clip=true]{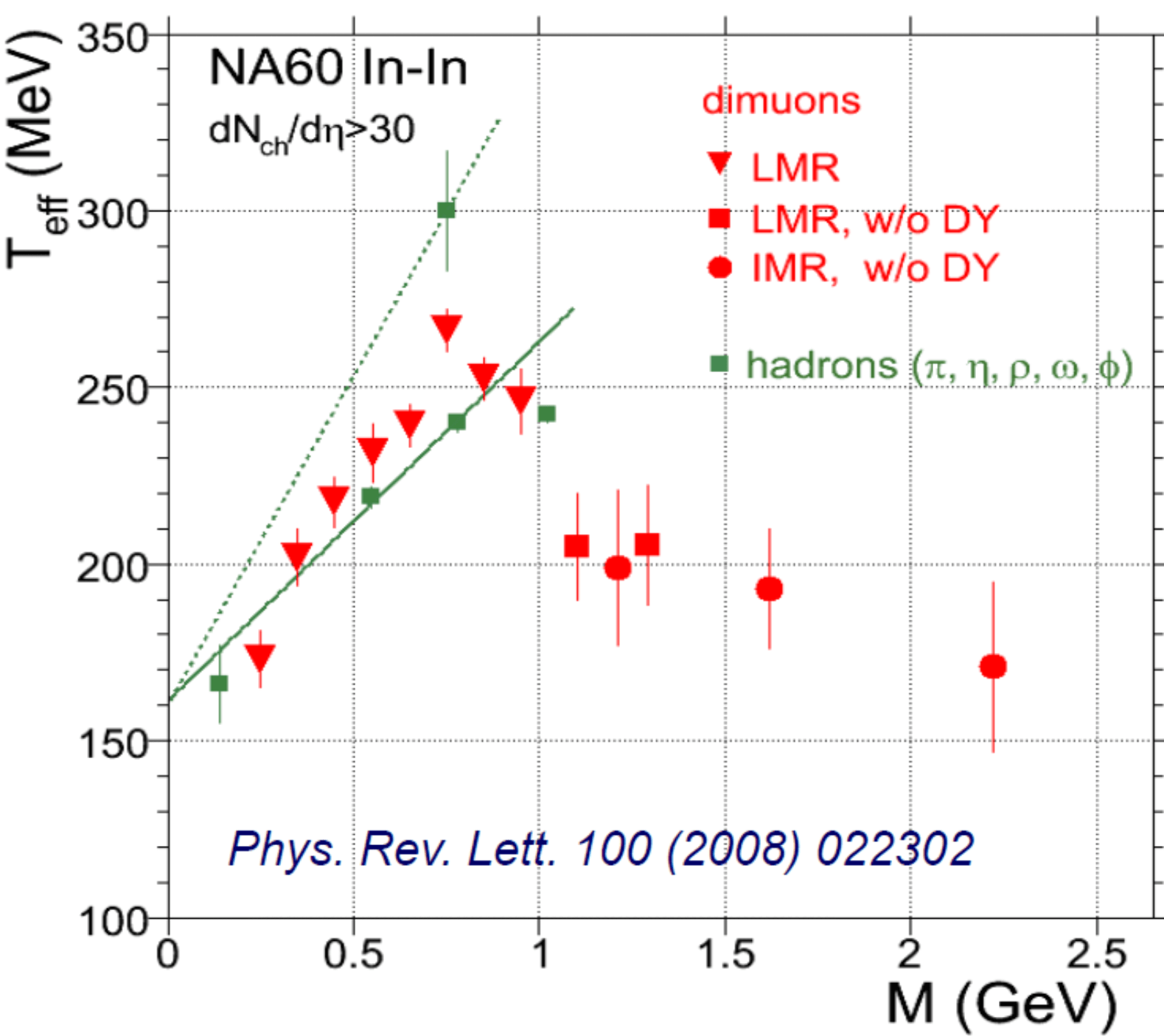}
\caption{
\textit{Left:} Direct photon signal observed in central Pb+Pb collisions~\cite{Manko:1999tgz} (WA98).
\textit{Right:} Effective temperature of directly produced dimouns in In+In collisions as function of dimuon mass~\cite{Arnaldi:2008er} (NA60). Top CERN SPS energy.
 }
\label{fig:q_signalsb}
\end{figure}

\vspace{0.2cm}
\textbf{The QGP discovery.}
Results on central collisions of medium size and heavy nuclei from the QGP search programme at the CERN SPS appeared to be consistent with the predictions for the QGP:
\begin{enumerate}[(i)]
     \item
    The relative yield of $J/\psi$ mesons is significantly suppressed compared to that in p+p and p+A interactions
    (NA38~\cite{Baglin:1990iv}, NA50~\cite{Alessandro:2004ap}, NA60~\cite{Arnaldi:2007zz}) as expected for the $J/\psi$ melting in a QGP
    (see Fig.~\ref{fig:q_signalsa} (\textit{left})).
    \item
    The relative strangeness yield is consistent with the yield expected for the equilibrium QGP. Moreover, it is significantly enhanced compared to that in p+p and p+A 
    interactions (see Fig.~\ref{fig:q_signalsa} (\textit{center}))
    (NA35~\cite{Bartke:1990cn}, NA49~\cite{Appelshauser:1998vn}). 
    Even larger enhancement is measured for the relative yield of multi-strange hyperons (see Fig.~\ref{fig:q_signalsa} (\textit{right})),
    (WA97~\cite{Andersen:1999ym}, NA57~\cite{Antinori:2006ij}). 
    Note that QGP formation was not expected in p+p and p+A collisions.
    \item
    Spectra of directly produced dimouns (virtual photons) and photons emerge above a dominant background at large mass respectively transverse momentum and show a thermal contribution with an effective temperature of about 200~\MeV. This is significantly larger than the expected transition temperature to QGP (see Fig.~\ref{fig:q_signalsb})
    (NA60~\cite{Arnaldi:2008er} and WA98~\cite{Manko:1999tgz}).
    \end{enumerate}

\vspace{0.2cm}
\textbf{Standard model of heavy-ion collisions.}
These and other results
established the standard picture of heavy-ion collisions~\cite{Florkowski:2010zz}:
\begin{enumerate}[(i)]
    \item 
    High density strongly interacting matter is created at the early stage of heavy-ion collisions. Starting at SPS collision energies it is in the QGP phase.
    \item
    The high-density matter enters a hydrodynamic expansion, cools down and emits photons and dileptons.
    \item
    At the phase transition temperature, $T_C \approx 150~\MeV$, hadrons are created.
    Statistical haronization models fit hadron yields at this stage quite well.
    \item 
    The hadronic matter after hadronization is still dense enough to modify the hadron composition and continue expansion.
    \item
    At sufficiently low densities the hadron interaction rate drops to zero (freeze-out).
    resonances decay and long-lived hadrons freely fly away e.g. towards particle detectors at the CERN SPS.
\end{enumerate}

\vspace{0.2cm}
\textbf{Conclusion from the QGP-search.}
These major achievements were compiled by the heavy-ion community~\cite{Heinz:2000bk} and led to the CERN press release - 
on February 10, 2000 the CERN Director General Luciano Maiani said:
\textit{
The combined data coming from the seven experiments on CERN's Heavy Ion programme have given a clear picture of a new state of matter. This result verifies an important prediction of the present theory of fundamental forces between quarks. It is also an important step forward in the understanding of the early evolution of the universe. We now have evidence of a new state of matter where quarks and gluons are not confined. There is still an entirely new territory to be explored concerning the physical properties of quark-gluon matter. The challenge now passes to the Relativistic Heavy Ion Collider at the Brookhaven National Laboratory and later to CERN's Large Hadron Collider.}

This was in fact the moment when the majority of heavy-ion physicists moved to
study heavy-ion collisions at much higher energies at the Relativistic Heavy Ion Collider (RHIC) of Brookhaven National Laboratory (BNL).
Rich and precise results obtained during the period of 2000-2010 at RHIC provided extensive information on the properties of the QGP. There were already no doubts about QGP formation at the early stage of A+A collisions at the CERN SPS,  and all the more  at RHIC energies.  Two basic properties of the QGP were established at the RHIC BNL: jet quenching (deceleration  of high momentum partons in the hot QGP)  and a small ratio of  the shear viscosity $\eta$ to the entropy density $s$. It was estimated that $\eta/s\cong 0.1$, i.e the QGP appears to be an almost perfect liquid (see Ref.~\cite{Adams:2005dq,Adcox:2004mh,Back:2004je,Arsene:2004fa} for details).  

The situation after the announcement of the QGP discovery in 2000 at CERN was however rather confusing. Many were pretty sure about its formation in central Pb+Pb collisions at the CERN SPS, but unambiguous evidence of the QGP state was still missing. 
Needless to say that the Nobel prize for the QGP discovery was not yet awarded.  
This may be attributed to the difficulty of obtaining unique and quantitative predictions of
the expected QGP signals from QCD.

\vspace{0.2cm}
\textbf{Question marks.}
Let us briefly discuss questions addressed to the two main signals of the QGP: the $J/\psi$ suppression and the strangeness enhancement.

{\bf The $J/\psi $ suppression.} The standard picture of $J/\psi$  production in collisions of hadrons and nuclei assumes
a two step process: 
the creation of a $c\overline{c}$ pair in hard parton collisions at the very early stage of the reaction and a subsequent formation of a bound charmonium state or two open charm hadrons. Further more it was assumed that the initial yield of $c\overline{c}$ pairs is proportional to the yield Drell-Yan pairs. Then the $J/\psi $/(Drell-Yan pairs) ratio is expected to be the same in p+p, p+A and A+A collisions providing there are no other processes which can lead to $J/\psi$
disintegration and/or creation.
The measured suppression of the ratio in p+A collisions respectively to p+p interactions was interpreted as due to $J/\psi$ interactions with nucleons of the target nucleus and with hadronic secondaries (`co-movers'). 
In central Pb+Pb collisions at 158\AGeV the suppression was found to be significantly stronger than expected in the models including nuclear and co-mover suppression. This anomalous $J/\psi$  suppression was interpreted as the evidence of the QGP creation in central Pb+Pb collisions at the top CERN SPS. 
However, the uncertainties related to the assumption $c\overline{c} \sim \textrm{Drell-Yan pairs}$ and estimates of the nuclear and co-mover suppression
lead to uncertainty in interpretation of the anomalous $J/\psi$  suppression as the QGP signal.
Moreover, models of  $J/\psi$  production in the later stages of the collision process have been developed:
\begin{enumerate}[(i)]
\item
the statistical model of $J/\psi$ production at the hadronization~\cite{Gazdzicki:1999rk},
\item
the dynamical and statistical models of $J/\psi$ production via coalescence of $c\overline{c}$ quarks~\mbox{\cite{Thews:2000rj,Levai:2000ne,BraunMunzinger:2000px,Gorenstein:2000ck}}.
\end{enumerate}
Clearly, in order to distinguish between different effects and verify the $J/\psi$ signal of the QGP creation systematic data on open charm production is needed.

\vspace{0.2cm}
\textbf{Strangeness enhancement.}
A fast equilibration of $s\overline{s}$ pairs was predicted as a QGP signature~\cite{Rafelski:1982pu,Rafelski:2019twp}. This is mainly because of the small mass of the strange quark, $m_s\sim 100$~\MeV compared to the QGP temperature: $T \ge T_C > m_s \approx 100$~\MeV. The estimated strangeness equilibration time was found to similar to the life time of the QGP phase in heavy-ion collisions at high energies.
In fact the strangeness yield measured in A+A collisions at the top SPS energy and above corresponds to the QGP equilibrium yield, for recent review see Ref.~\cite{Rafelski:2019twp}.
Moreover, it was estimated that the strangeness equilibration time in the confined matter is about 10 times longer than the life time of the hadronic phase in A+A collisions. This is because masses of strange hadrons, starting form the lightest one, the kaon ($m_K\sim 500$~\MeV) are much larger than the maximum temperature of the hadron-resonance gas $T\le T_C \approx 150$~\MeV.  Thus a small yield of strangeness was expected for reactions in which the QGP was not expected, p+p and p+A interactions and A+A collisions at low energies. Consequently the enhanced production of strangeness was predicted as the next QGP signal~\cite{Glendenning:1984ta}.
 
The strangeness enhancement is quantified by comparing a strange-hadron to pion ratio in A+A collisions with that in p+p interactions.
In particular a double ratio is calculated:
\eq{\label{str-pion}
R(\sqrt{s_{NN}})= \frac{\langle K^+\rangle_{AA}/\langle \pi^+\rangle_{AA}}{\langle K^+\rangle_{pp}/\langle \pi^+\rangle_{pp}}~,
}
where $\langle\ldots \rangle_{AA}$ and $\langle \ldots \rangle_{pp}$ denote the event averages of $K^+$ and $\pi^+$ yields in, respectively, A+A collisions and p+p interactions at the same center of mass collision energy $\sqrt{s_{NN}}$ of the nucleon pair.  Ratios of different strange hadrons to pions  
were considered, e.g.,  $\langle K+\overline{K}\rangle/\langle \pi\rangle$, $\langle \Lambda\rangle/\langle \pi\rangle$, $\ldots$, $\langle\Omega\rangle /\langle \pi \rangle $,  
and then analyzed by forming the double ratios $R$, similar to the one given by Eq.~(\ref{str-pion}).
 The confrontation of this expectation with the data was for the
first time possible in 1988 when results from the SPS and the AGS
became available. NA35  reported~\cite{Bartke:1990cn} that in central S+S collisions at 200\AGeV
the strangeness to pion ratio (\ref{str-pion}) is indeed about two times higher than in nucleon-nucleon interactions at
the same energy per nucleon. But an even larger enhancement ($R= 14 - 5 $) 
was measured at the AGS at 2$A -10A$~GeV~\cite{Pinkenburg:2001fj,Abbott:1990mm} demonstrating that strangeness enhancement {\it increases} with {\it decreasing} collision energy.
Moreover, the enhancement factor (\ref{str-pion}) should evidently go to infinity at the threshold energy of strange hadron production in nucleon-nucleon interactions. 
Note also that the strangeness neutrality introduced to statistical models using the canonical ensemble leads to a suppression of the relative yield of strange particles in systems with a low multiplicity of strangeness carriers~\cite{Rafelski:1980gk}, e.g. p+p interactions at SPS energies.  
In any case, the AGS measurements indicating a strangeness enhancement larger than that at the CERN SPS show clearly difficulties in interpreting the strangeness enhancement as the QGP signal. 

\vspace{0.2cm}
\textbf{New strategy.}
Difficulties in interpretation of the QGP signatures forced scientists to rethink the QGP-hunt strategy. 
The emerging new strategy was similar to the one followed by physics studying molecular liquids and gases. In these essentially simpler and familiar cases it is also sometimes  difficult to distinguish the properties of a dense gas from those of a liquid. It is much easier to identify the effects of the liquid-gas transition.  
Thus, if one believes that the QGP is formed in central Pb+Pb collisions at the top SPS energy one should observe qualitative signals of the {\it transition} to the QGP at a lower collision energy. 
Several such signals were predicted within the statistical model of the early stage
~\cite{Gazdzicki:1998vd}.
Their observation would  serve as strong evidence of QGP creation in heavy-ion collisions at high enough collision energies.

This idea  motivated some of us to propose the collision energy scan at the CERN SPS with the aim to search for the {\it onset of deconfinement}. This was the beginning of the search for the critical structures in heavy-ion collisions, for detail see the next section and Ref.~\cite{Gazdzicki:2003fj}.

\section{ Critical structures }
\label{sec:cs}


\subsection{Evidence for the onset of deconfinement}
\label{sec:cs_ood}


\textbf{Predicted signals of the onset of deconfinement.}
The experimental search for the onset of deconfinement in heavy ion collisions at the CERN SPS was shaped by several model predictions of possible measurable signals:
\begin{enumerate}[(i)]
    \item 
    characteristic enhanced production of pions and suppression of the strangeness to pion ratio ~\cite{Gazdzicki:1998vd},
    \item
    softening of collective flow of hadrons~\cite{Gorenstein:2003cu,Csernai:1999nf,Bleicher:2000sx,Bleicher:2005tb}, which should be observed in hadron distributions in transverse~\cite{Gorenstein:2003cu} and longitudinal momenta~\cite{Bleicher:2005tb} and azimuth angle~\cite{Csernai:1999nf,Bleicher:2000sx}.
\end{enumerate}

\vspace{0.2cm}
\textbf{Measurements at the CERN SPS and RHIC BES.}
The search for the onset of deconfinement at the CERN SPS started 
in 1999 with the data taking on Pb+Pb collisions at 40\AGeV.
The data were registered by NA49, NA45, NA50 and NA57. 
In 2000 a beam at 80\AGeV was delivered to NA49 and NA45.
The program was completed in 2002 by runs of NA49 and NA60 at 20\AGeV and 30\AGeV.
Thus, together with the previously recorded data at 158\AGeV, NA49 gathered data at  five collision energies. Other experiments collected data at two (NA50, NA57) or three (NA45, NA60) energies. Starting in 2010 the beam energy scan program BES was started at RHIC with the aim of covering the low
energy range overlapping with the CERN SPS and providing important
consistency checks on the measurements.

\begin{figure}[hbt]
\centering
 \includegraphics[width=0.99\linewidth,clip=true]{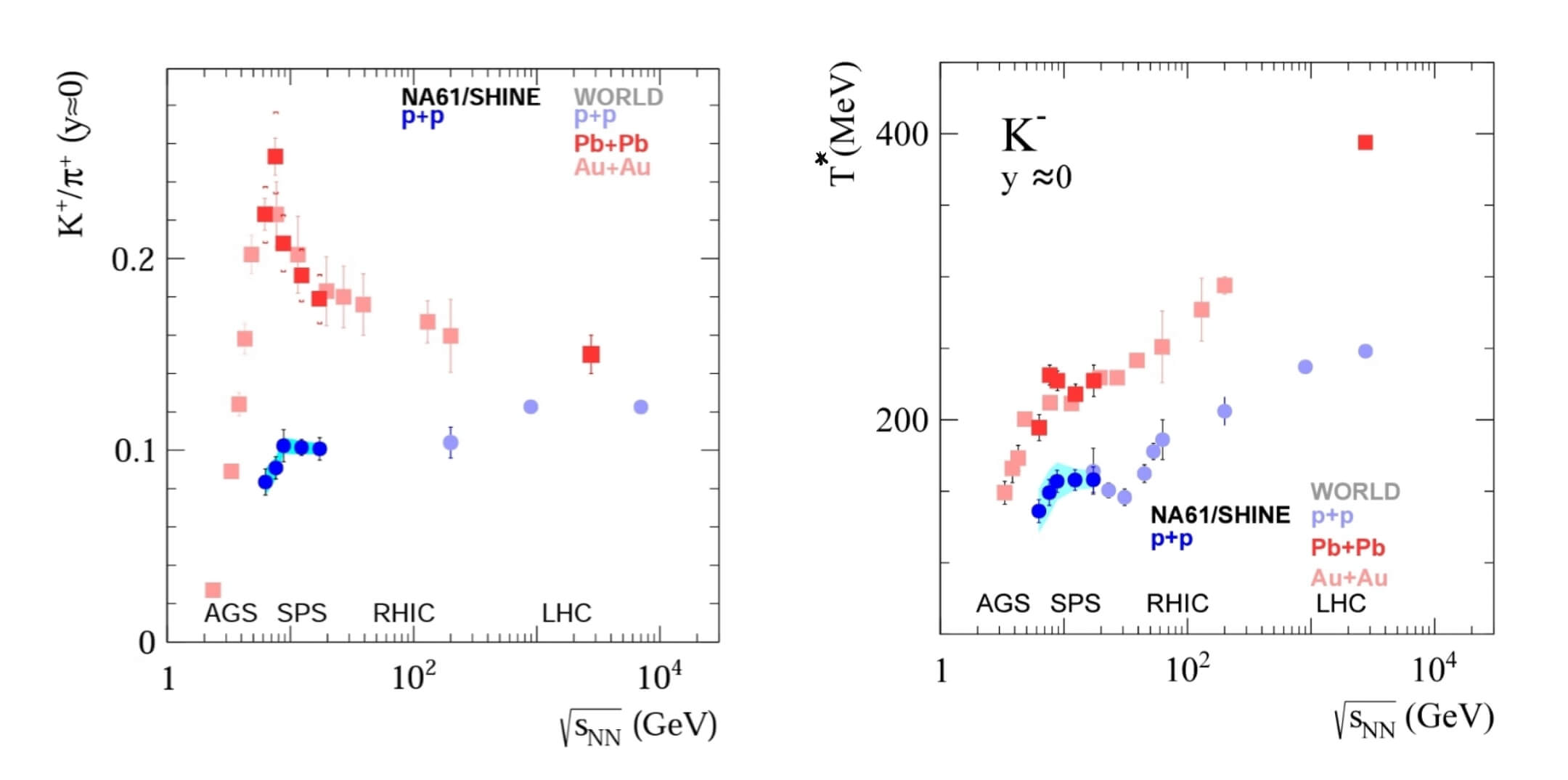}
\caption{
Examples of results illustrating the observation of the onset-of-deconfinement signals in central Pb+Pb (Au+Au) collisions~\cite{Aduszkiewicz:2019zsv}, see text for details and more references.
 }
\label{fig:cs_ood_signals}
\end{figure}

\vspace{0.2cm}
\textbf{Discovery of the onset of deconfinement.}
Results on the collision energy dependence of hadron production in central Pb+Pb collisions from the onset-of-deconfinement search programme at the CERN SPS~\cite{Afanasiev:2002mx,Alt:2007aa} appeared to be consistent with the predicted signals (for review see Ref.~\cite{Gazdzicki:2010iv}):
\begin{enumerate}[(i)]
\item
The average number of pions per wounded nucleon, $\langle N_\pi \rangle /\langle W \rangle $, 
in low energy A+A collisions is smaller that this value in p+p reactions.  This relation is however changed to the opposite at  collision energies larger than $\approx$~30\AGeV, the so-called {\bf kink} structure. 
     \item
    The collision energy dependence of the $\langle K^+\rangle_{AA}/\langle \pi^+\rangle_{AA}$ ratio shows the so-called \textbf{horn} structure. Following a fast rise the ratio passes through a maximum in the SPS range, at approximately 30\AGeV, and then decreases and settles to a plateau value at higher energies. This plateau was found to continue up to the RHIC and LHC energies.
    \item
    The collision energy dependence of the inverse slope parameter of the transverse mass spectra, $T^*$, of charged kaons shows the so-called \textbf{step} structure. Following a fast rise the $T^*$ parameter passes through a stationary region (or even a weak minimum for $K^-$), which starts at the low SPS energies, approximately 30\AGeV, and then enters a domain of a steady increase above the top SPS energy.
    \end{enumerate}
    
Figure~\ref{fig:cs_ood_signals} shows examples of the most recent plots~\cite{Aduszkiewicz:2019zsv} illustrating the observation of the onset-of-deconfinement signals. As seen data from the RHIC BES~I programme
(2010-2014) and LHC (see Ref.~\cite{Aduszkiewicz:2019zsv} for references to original experimental papers) confirm the NA49 results and their interpretation.

Two comments are appropriate here.
The strangeness to pion ratio, e.g., 
$\langle K^+\rangle_{AA}/\langle \pi^+\rangle_{AA}$,
strongly increases with collision energy in the hadron phase. This happens because 
$m_K/T\gg 1$, whereas $m_\pi/T\cong 1$.
Thus, a much stronger increase with increasing temperature is expected for the multiplicities of heavy strange hadrons than that of pions. The strangeness to pion ratio reaches its maximum inside the hadron phase at the onset of the deconfinement. 
The plateau-like behavior  at high collision energies reflects the approximately constant value  of the 
strangeness to entropy ratio in the QGP.  It equals to the ratio of the degeneracy factor of strange quarks, 
\eq{\label{s-quark}
g_s=\frac{7}{8}\cdot 2\cdot 2\cdot 3 =10.5~,
}
to the total degeneracy factor of the quark-gluon constituents in the QGP,
\eq{\label{qg}
g= 2\cdot 8+ \frac{7}{8}\cdot 2\cdot 2\cdot 3 \cdot 3  = 47.5~.
}
These degeneracy factors count 2 spin states of quarks and gluons, 3 flavor quark states, 8 colour states of gluons and 3 colour states of quarks, one more factor 2 appears due to antiquarks (the factor 7/8 is due to the Fermi statistics of quarks).   
The strangeness to entropy ratio in the HRG at the largest  hadron temperature
$T\cong T_H\cong 150$~MeV appears to be larger than this ratio in the QGP which is approximately constant at all QGP temperatures $T\ge T_H$. Therefore, 
the transition region from hadron matter to the QGP 
reveals itself as the {\it suppression} of strangeness yield relative to pion yield.

The second comment concerns the inverse slope parameter $T^*$ of the transverse mass ($m_T=\sqrt{m^2+p_T^2}$) spectrum 
\eq{\label{pT}
\frac{dN}{m_T dm_T}\sim \exp\left(-\,\frac{m_T}{T^*}\right)~.
}
The parameter $T^*$ is sensitive to both the thermal and collective motion transverse to the collision axis and behaves as
\eq{\label{T*}
T^*\cong T+ \frac{1}{2}m v_T^2~,
}
where $T$ is the temperature and $v_T$ is the transverse collective (hydrodynamic) velocity of the hadronic matter at the kinetic freeze-out. The parameter $T^*$ increases strongly with collision energy 
up to the energy $\approx$~30\AGeV. This is because an increasing collision energy leads to an increase of both terms in Eq.~(\ref{T*}) - the temperature  $T$ and velocity $v_T$ - in the hadron phase ($v_T$ increases due to the increase of the pressure). At collision energy larger than $\approx$~30\AGeV the parameter $T^*$ is approximately independent of the collision energy in the SPS energy range. In this region one expects the transition between confined and deconfined matter. In the transition region both values - $T$ and $v_T$ - remain approximately constant, and this leads to the plateau-like structure in the energy dependence of the $T^*$ parameter. 
At  RHIC--LHC energies, the parameter $T^*$ again increases with collision energy. The early stage QGP pressure increases with collision energy, and thus $v_T$ in Eq.~(\ref{T*}) increases too. 

\vspace{0.1cm}
The workshop \textit{Tracing the onset of deconfinement in nucleus-nucleus collisions}, ECT* Trento, April 24-29, 2004, 
summarized the results from the energy scan programme at the CERN SPS and concluded that future measurements in the SPS energy range are needed~\cite{Gazdzicki:2004gss}.
The goal is  to search for the deconfinement critical point and study system size dependence of the onset of deconfinement. Possibilities to perform  these measurements at the CERN SPS, FAIR SIS300 and RHIC were discussed. The event initiated a series of the \textit{Critical point and onset of deconfinement} workshops.

\begin{figure}[hbt]
\centering
 \includegraphics[width=0.45\linewidth,clip=true]{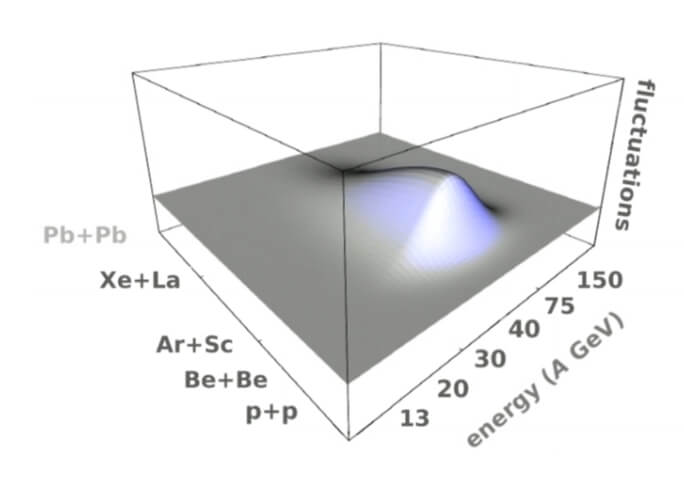}
 \includegraphics[width=0.45\linewidth,clip=true]{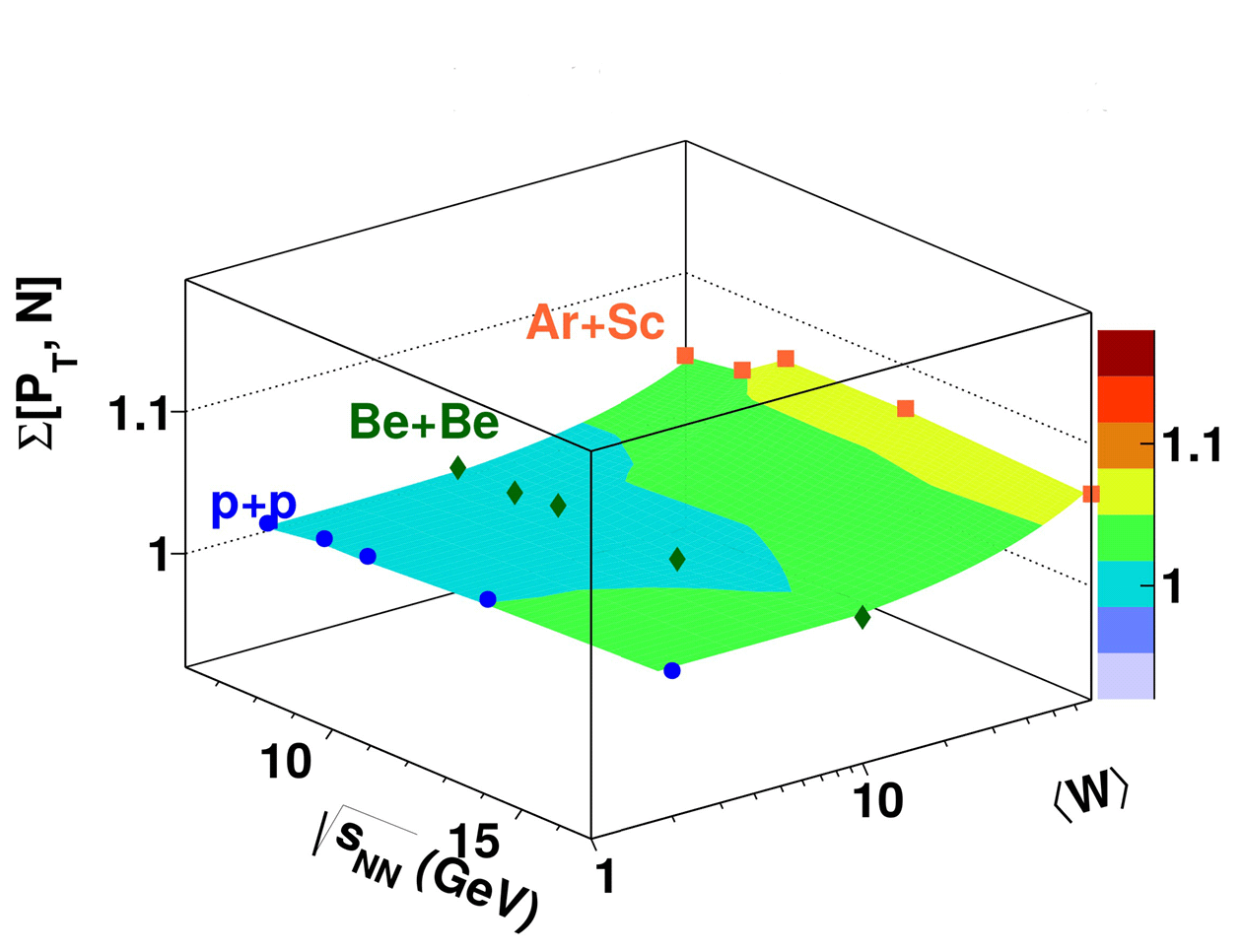}
\caption{
 \textit{Left:}
 Sketch of the expected signal of the deconfinement critical point - a maximum of fluctuations in the (nuclear mass number)-(collision energy) plane.
 \textit{Right:}
     Results from the \NASixtyOne two dimensional scan of energy and system size for (pion multiplicity)-(transverse momentum) fluctuations in terms of the strongly intensive quantity $\Sigma[P_T,N]$~\cite{Gazdzicki:2017zrq}.
 }
\label{fig:cs_cp_signals}
\end{figure}

\subsection{Searching for deconfinement critical point}
\label{sec:cs_cp}

\textbf{Predicted d-CP signals.}
The possible existence and location of the deconfinement critical point (d-CP) is a subject of
vivid theoretical discussion, for a recent review see Ref.~\cite{Bzdak:2019pkr}. 
The experimental search for the d-CP
in A+A collisions at the CERN SPS was shaped by several model predictions (for detail see Ref.~\cite{Bialas:1990xd}) of its potential signals:
\begin{enumerate}[(i)]
    \item 
    characteristic multiplicity fluctuations of hadrons~\cite{Bialas:1990xd,Antoniou:2000ms,Hatta:2003wn,Antoniou:2006zb},
    \item
    enhanced fluctuations of (pion multiplicity)-(transverse momentum)~\cite{Stephanov:1999zu},
\end{enumerate}

The signals were expected to have a maximum in the parameter space of collision energy and nuclear mass number of colliding nuclei - \textbf{the hill of fluctuations}~\cite{Gazdzicki:2006fy}.
This motivated \NASixtyOne to perform a two dimensional scan at the CERN SPS~\cite{Gazdzicki:995681} in these two parameters, which are well controlled in laboratory experiments.

\begin{figure}[hbt]
\centering

 \includegraphics[width=0.99\linewidth,clip=true]{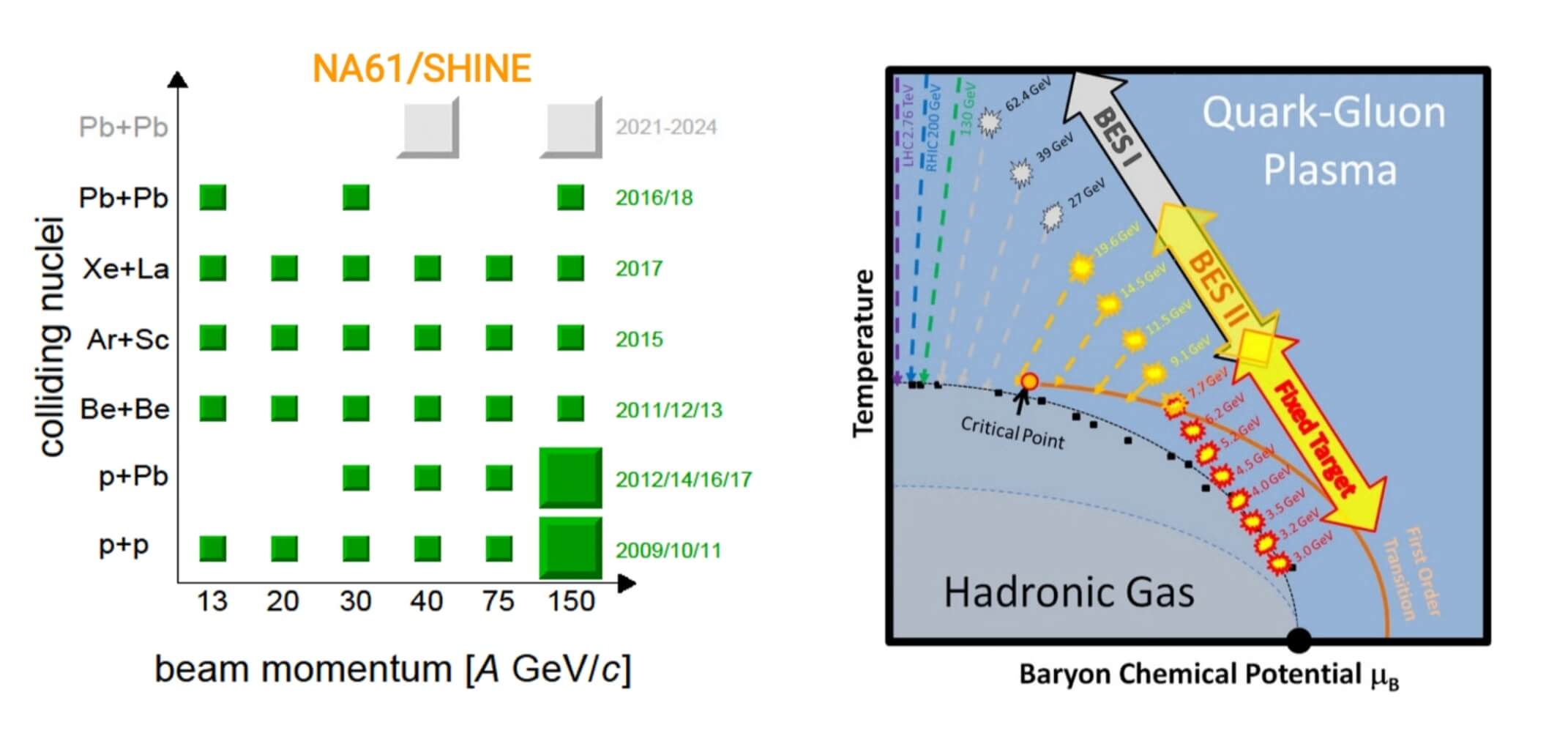}

\caption{
 Summary of data recorded by \NASixtyOne at the CERN SPS~(\textit{left}) and STAR at RHIC~(\textit{right}) relevant for the search for the deconfinement-CP, see text for details. 
 }
\label{fig:cs_cp_exp}
\end{figure}

\vspace{0.2cm}
\textbf{Measurements at SPS and RHIC.}
The systematic search for the d-CP of strongly interacting matter was started in 2009 with the \NASixtyOne data taking on p+p interactions at six beam momenta in the range from 13\AGeVc to 158\AGeVc.
In the following years data on Be+Be, Ar+Sc, Xe+La and Pb+Pb collisions were
recorded, see Fig.~\ref{fig:cs_cp_exp}~(\textit{left}) for an overview.

In 2010 the beam energy scan (BES-I and BES-II) with Au+Au collisions started at the BNL RHIC~\cite{Odyniec:2019kfh}. Search for the deconfinement critical point has been the most important goal of this programme. Above the collision energy of $\sqrt{s_{NN}} = 7.7~\GeV$ ($\approx$~30\AGeVc) the scan was conducted in the collider mode, whereas below in the fixed target mode. The location of the recorded data in the phase diagram are shown in Fig.~\ref{fig:cs_cp_exp}~(\textit{right}).

\vspace{0.2cm}
\textbf{Status of the d-CP search.}
Many experimental results have already been obtained within the d-CP search programmes at SPS and RHIC, for a recent review see Ref.~\cite{Czopowicz:2020twk}. Five of them were considered as possible indications of the d-CP and are presented and discussed in the following.

\begin{enumerate}[(i)]
\item
A maximum of fluctuations is expected in a scan of the phase diagram (see Fig.~\ref{fig:cs_cp_signals} (\textit{left})). Measurements of (pion multiplicity)-(transverse momentum) fluctuations from \NASixtyOne shown in Fig.~\ref{fig:cs_cp_signals} (\textit{right}) do not show such a 
feature~\cite{Gazdzicki:2017zrq}.
\item
The energy dependence of fluctuations of conserved quantities such as the net baryon number is predicted to be sensitive to the presence of the d-CP. This holds in particular for higher moments. The scaled third and fourth moments of the net-proton multiplicity distribution in Au+Au collisions from the STAR experiment is plotted in Fig.~\ref{fig:cs_cp_results1}~\cite{Adam:2020unf}. The non-monotonic behaviour of the fourth moment in central collisions and its sign change around $\sqrt{s_{NN}} \approx 7$~\GeV is debated as possible indication of the d-CP.

\begin{figure}[hbt]

\centering
 \includegraphics[width=0.7\linewidth,clip=true]{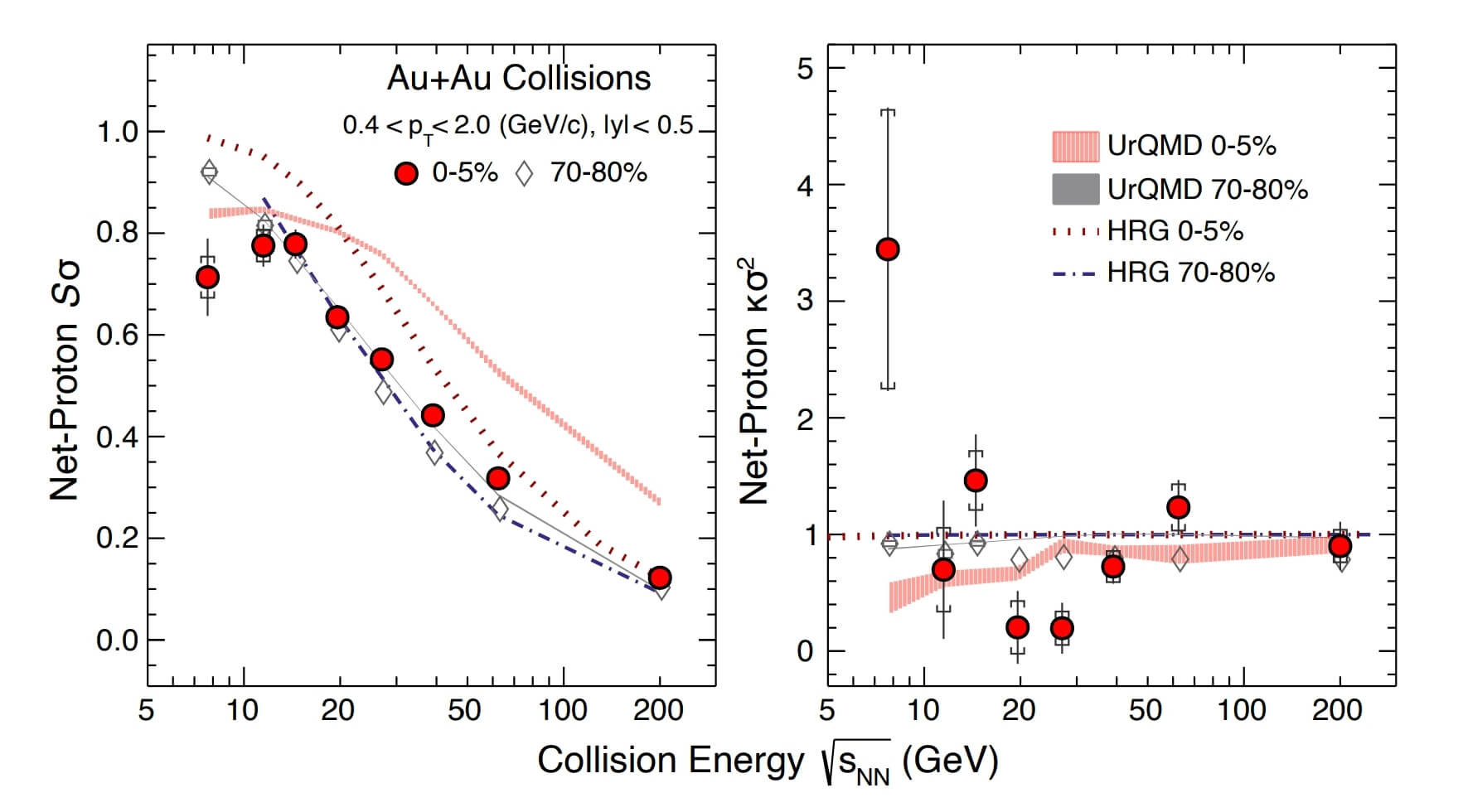}
\caption{
 The energy dependence of the scaled third (\textit{left}) and fourth (\textit{right}) moments of the net-proton multiplicity distribution in central and peripheral Au+Au collisions from the STAR experiment~~\cite{Adam:2020unf}.
 }
\label{fig:cs_cp_results1}
\end{figure}

\item
At the d-CP the correlation length diverges and leads to power-law type fluctuations of the baryon number. These were investigated by the \NASixtyOne experiment by measuring the momentum bin size dependence of the scaled second factorial moment of the proton multiplicity distribution (intermittency study) in semi-central Ar+Sc collisions at  $\sqrt{s_{NN}} \approx 17~\GeV$~\cite{Mackowiak:2019qm}. While previous measurements by the NA49 experiment in Si+Si collisions indicated a signal the new measurements 
shown in Fig.~\ref{fig:cs_cp_results2} do not confirm the effect.

\begin{figure}[hbt]

\centering
 \includegraphics[width=0.4\linewidth,clip=true]{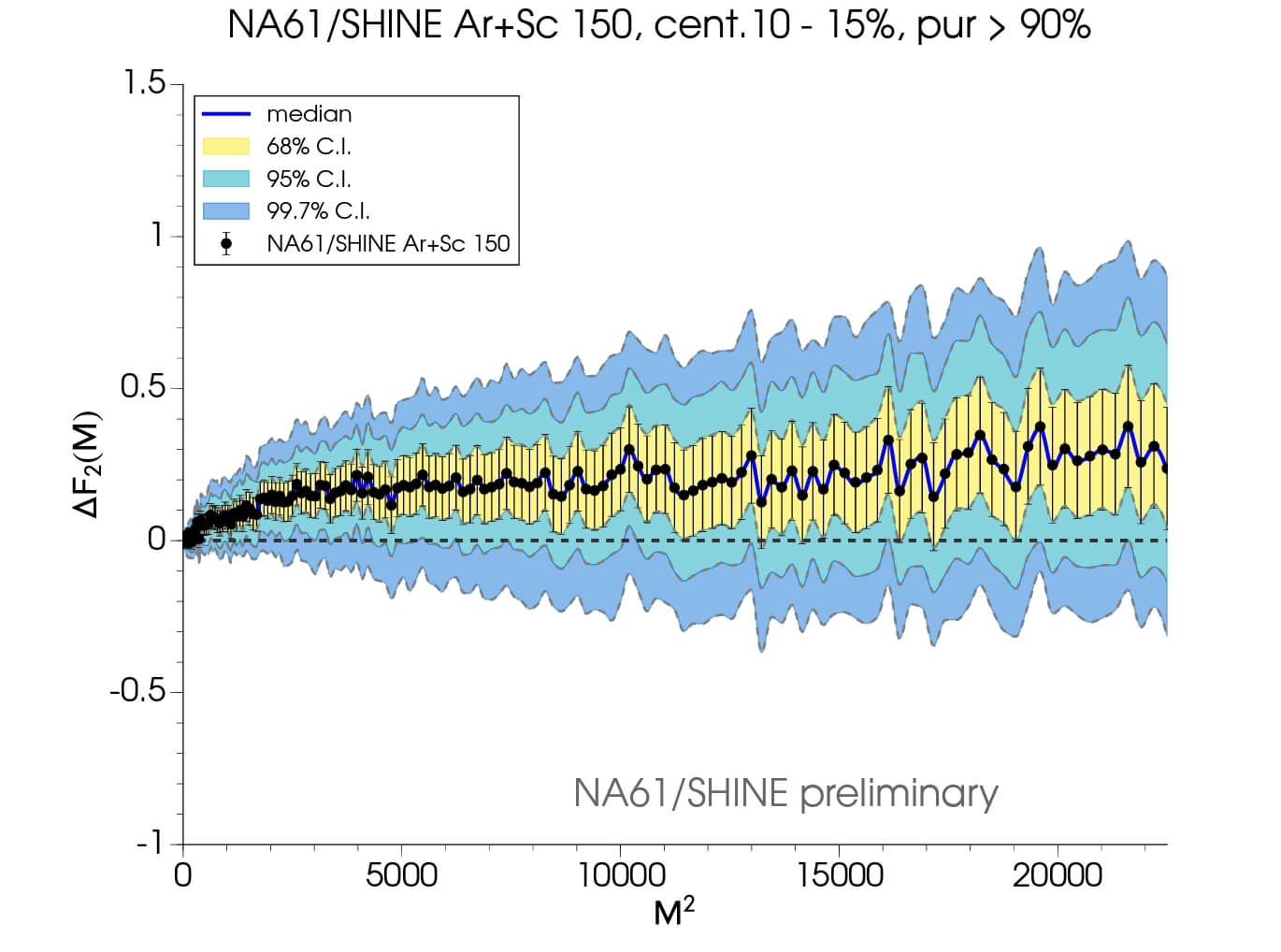}
\caption{
 Scaled second factorial moment $\Delta F_2$ (background subtracted) of the proton multiplicity distribution as function of the number of subdivisions M of transverse momentum space obtained in Ar+Sc collisions at $\sqrt{s_{NN}} \approx$ 17~\GeV~\cite{Mackowiak:2019qm}.
 }
\label{fig:cs_cp_results2}
\end{figure}

\item
The ratio of yields of light nuclei production can be related to nucleon number fluctuations~\cite{Liu:2019nii}. The measurements from STAR in central Au+Au collisions show strong collision energy dependence and peak at $\sqrt{s_{NN}} \approx 20 - 30~\GeV$~\cite{Xu:2019qm}. These results are presented in Fig.~\ref{fig:cs_cp_results3}. Such behaviour is not reproduced by model calculations without a d-CP and may thus be attributable
to a critical point.
\begin{figure}[hbt]

\centering
 \includegraphics[width=0.4\linewidth,clip=true]{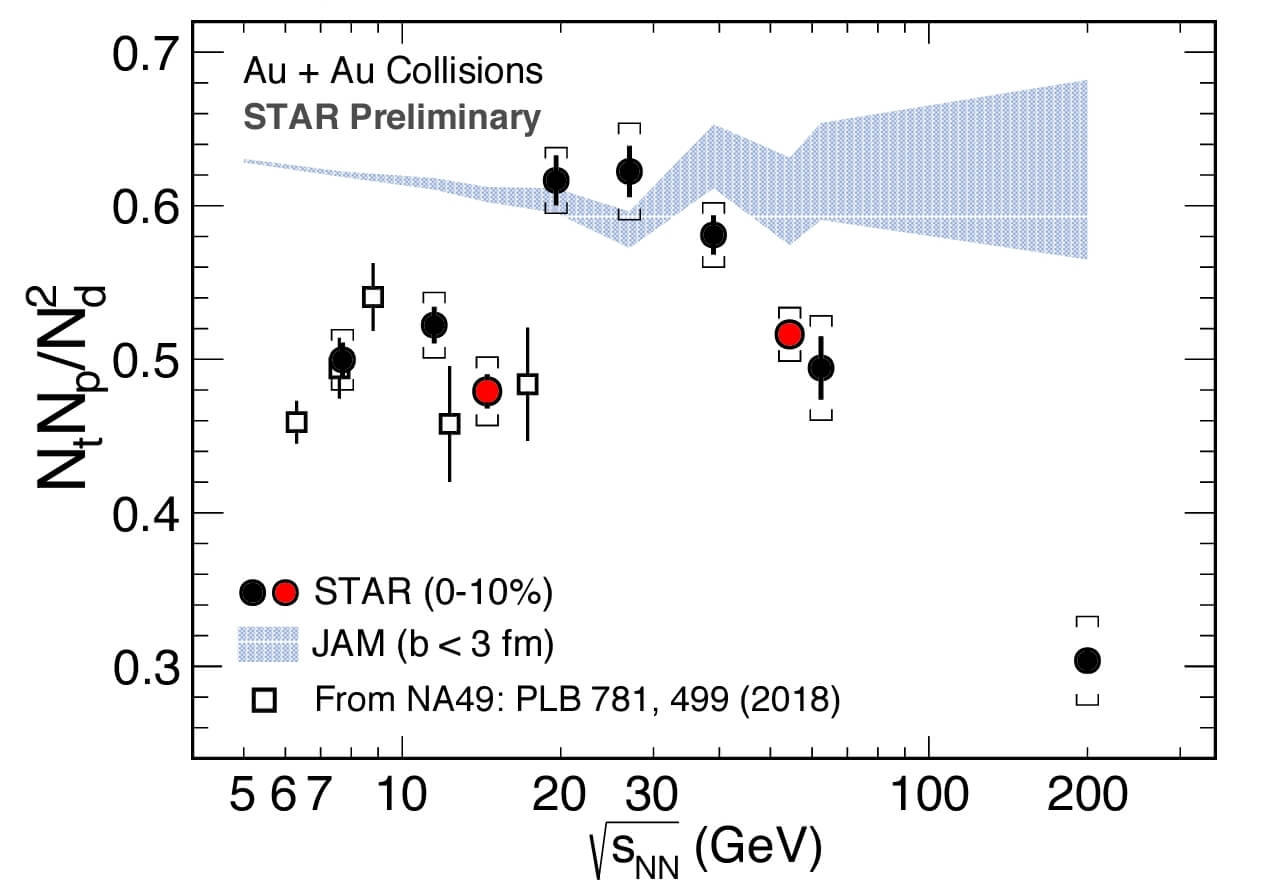}
\caption{
 Energy dependence of the ratio of yields of light nuclei production in central Au+Au collisions~\cite{Xu:2019qm} measured by the STAR experiment at RHIC. 
 }
\label{fig:cs_cp_results3}
\end{figure}

\item
The energy and centrality dependence of short-range two-pion correlations as parameterized by source radius parameters determined from Bose-Einstein correlation analysis was used to search for indications of the d-CP~\cite{Lacey:2014wqa,Adamczyk:2014mxp}.
The result for the difference $R^2_{out}$ - $R^2_{side}$ in Au+Au collisions at RHIC is shown in Fig.~\ref{fig:cs_cp_results4}.
A finite size scaling analysis of these results led to an estimate of the position of a d-CP at  $T \approx$~165~MeV and 
$\mu_B$ $\approx$ 95~MeV.

\begin{figure}[hbt]

\centering
 \includegraphics[width=0.4\linewidth,clip=true]{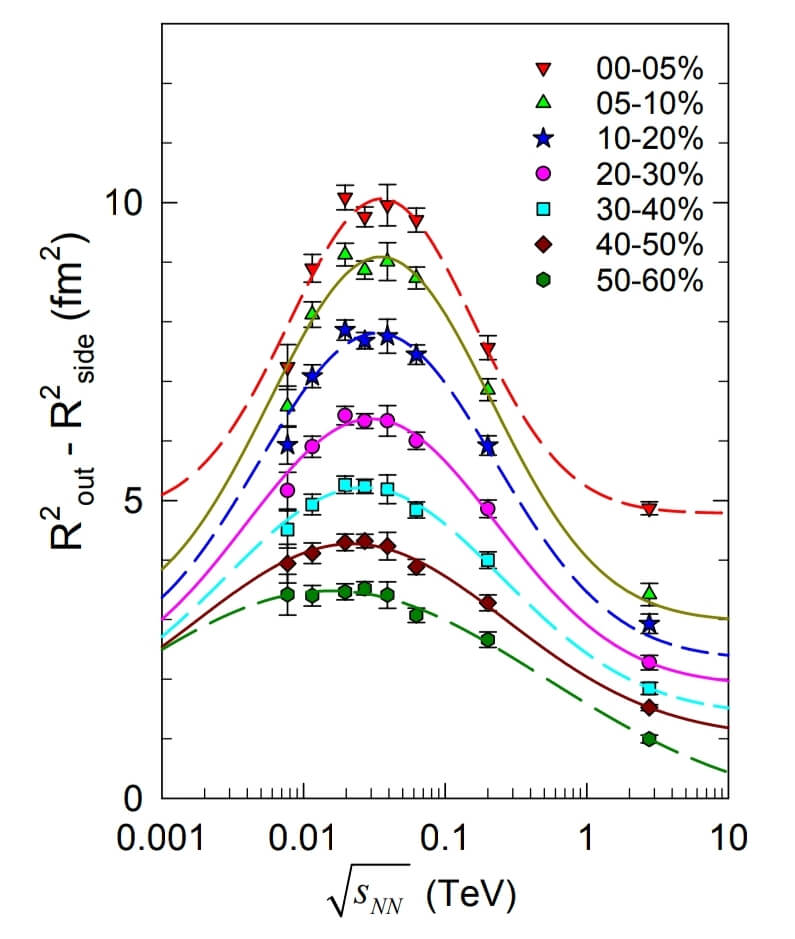}
\caption{
 Centrality and energy dependence of the difference $R^2_{out}$ - $R^2_{side}$
 of radius parameters obtained from Bose-Einstein two-pion correlation analysis in Au+Au collisions from the PHENIX experiment at RHIC
 ~\cite{Lacey:2014wqa,Adamczyk:2014mxp}.
 }
\label{fig:cs_cp_results4}
\end{figure}

\end{enumerate}

\begin{figure}[hbt]

\centering
 \includegraphics[width=0.6\linewidth,clip=true]{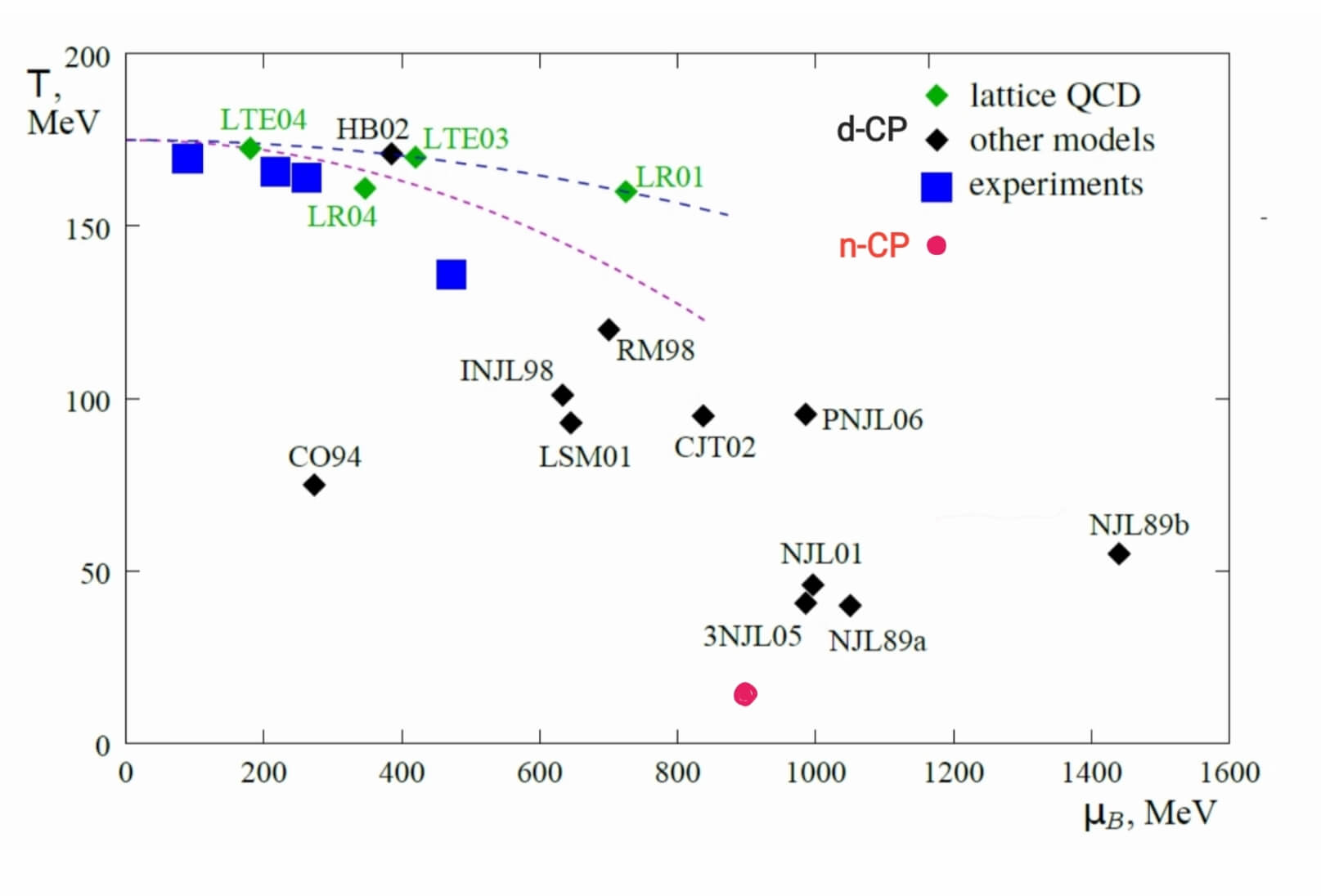}
\caption{
Compilation of theoretical predictions~\cite{Stephanov:2007fk} and experimental hints~\cite{Czopowicz:2020twk} on the location of the deconfinement critical
point, d-CP, in the phase diagram of strongly interacting matter. The position 
of nuclear critical point, n-CP, as suggested by theoretical and experimental results is indicated for comparision.
 }
\label{fig:cs_cp_results_pd}
\end{figure}

These observations, when interpreted as due to the d-CP, yield different estimates of the d-CP location on the
phase diagram of strongly interacting matter, see Fig.~\ref{fig:cs_cp_results_pd}~\cite{Czopowicz:2020twk}.
Thus, as for now, the experimental results concerning the d-CP are inconclusive. New results from \NASixtyOne and STAR BES-II are expected within the coming years. 
\vspace{0.2cm}

\textbf{Nuclear and deconfinement critical points.}
The nuclear critical point (n-CP) corresponds to the liquid-gas phase transition in the system of interacting nucleons and is located at small temperature $T_C\approx 19$~\MeV and large baryonic chemical potential $\mu_B\approx 915$~\MeV, see Fig.~\ref{fig:v_qgp} (\textit{middle}) and Fig.~\ref{fig:cs_cp_results_pd} for illustration.

The effect of the n-CP on fluctuations of conserved charges,
baryon number (B), electric charge (Q), and strangeness (S),
was studied in Refs.~\cite{Vovchenko:2017ayq, Poberezhnyuk:2019pxs} within the 
HRG
model with van der Waals interactions between baryons and between anti-baryons. 
The second, third, and fourth order cumulants 
(susceptibilities) are
calculated in the grand canonical ensemble from the pressure function by taking the derivatives over the corresponding chemical potentials:
\eq{
\chi^{i}_{n}& =\frac{\partial ^{n}\left( p/T^{4}\right) }{\partial \left( \mu_{i} 
/T\right) ^{n}}~,
}
where $i$ stands for $B, Q, S$ and $n$ is the moment order. 

The obtained results show that
the n-CP may significantly impact event-by-event fluctuations in A+A collisions even at high energies.  
Thus, the nuclear-CP should be taken into account in  future searches for the deconfinement-CP.

\subsection{Indication of the onset of fireball}
\label{sec:cs_oof}

\textbf{Predictions of reference models on system-size dependence.}
There are two models often used to obtain reference predictions concerning the system-size dependence of hadron production properties~\cite{Gazdzicki:2013lda} - the Wounded Nucleon Model (WNM)~\cite{Bialas:1976ed} and the Statistical Model (SM)~\cite{Rafelski:1980gk}. For the \kp/\pip ratio at the CERN SPS energies they read:  
\begin{enumerate}[(i)]
\item
The WNM prediction: the \kp/\pip ratio is independent of the system size (number of wounded nucleons).
\item
The SM prediction: in the canonical formulation incorporating global quantum number conservation the \kp/\pip ratio increases monotonically with the system size and approaches the limit given
by the grand canonical approximation of the model. The rate of this increase is the fastest for small systems.
\end{enumerate}

\begin{figure}[hbt]

\centering
 \includegraphics[width=0.47\linewidth,clip=true]{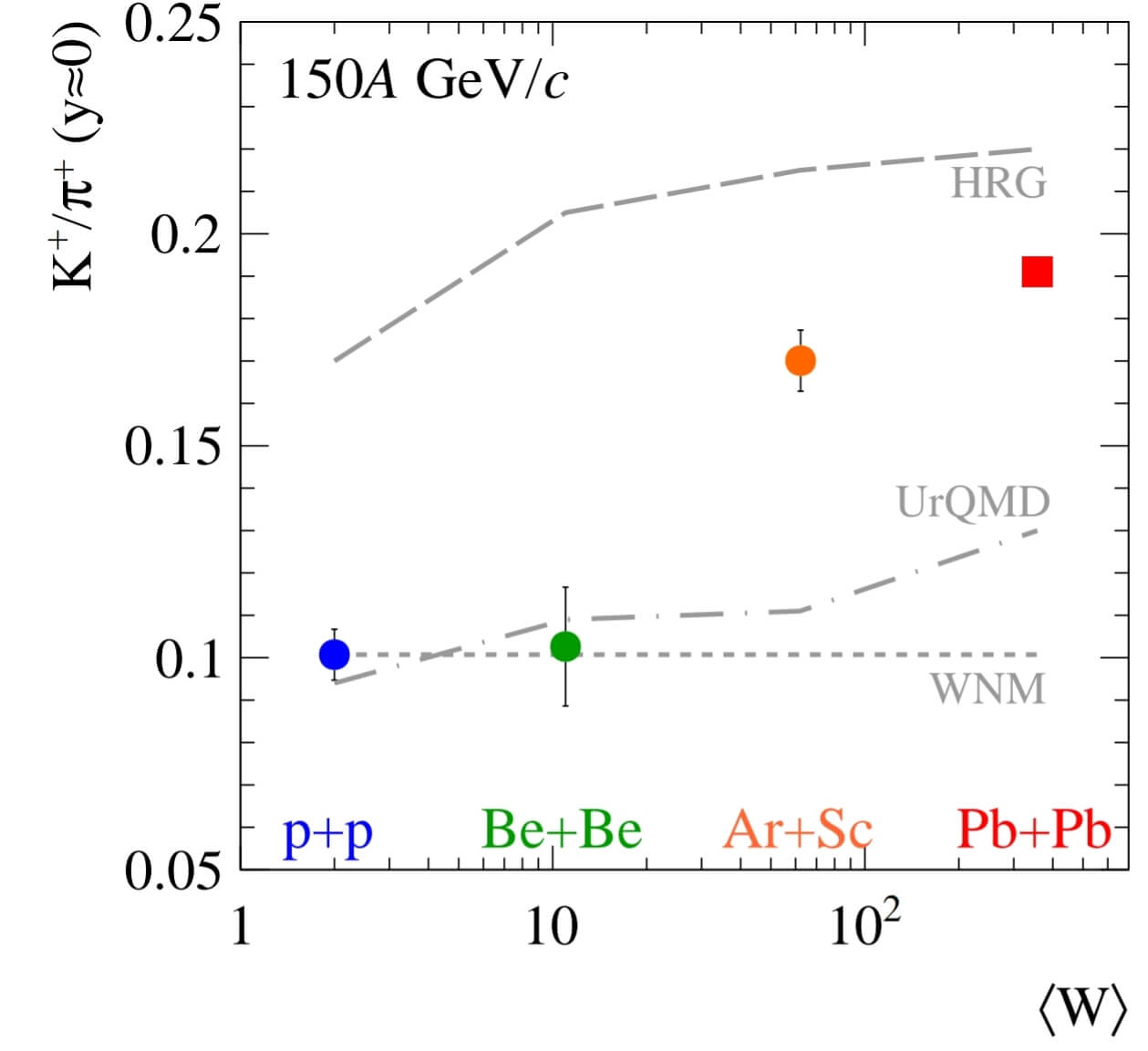}
 \includegraphics[width=0.45\linewidth,clip=true]{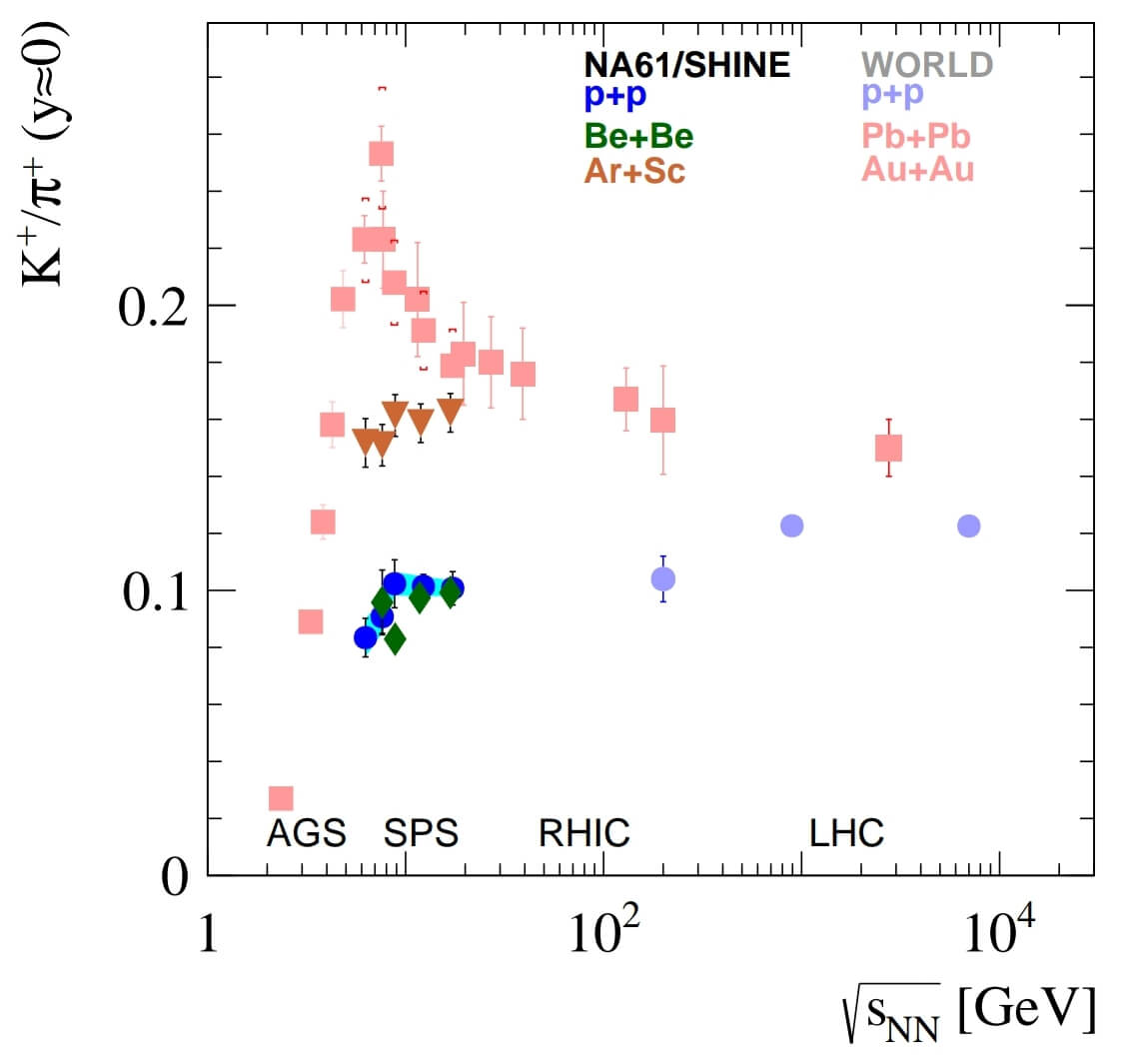}
\caption{
 Measurements of the the \kp/\pip ratio in p+p, Be+Be, Ar+Sc and Pb+Pb collisions: system size dependence at 150\AGeVc~\cite{Podlaski:2019} (\textit{left}) and  collision energy dependence~\cite{Aduszkiewicz:2019zsv} (\textit{right}). 
 }
\label{fig:cs_oof_results}
\end{figure}

\begin{figure}[hbt]

\centering
 \includegraphics[width=0.47\linewidth,clip=true]{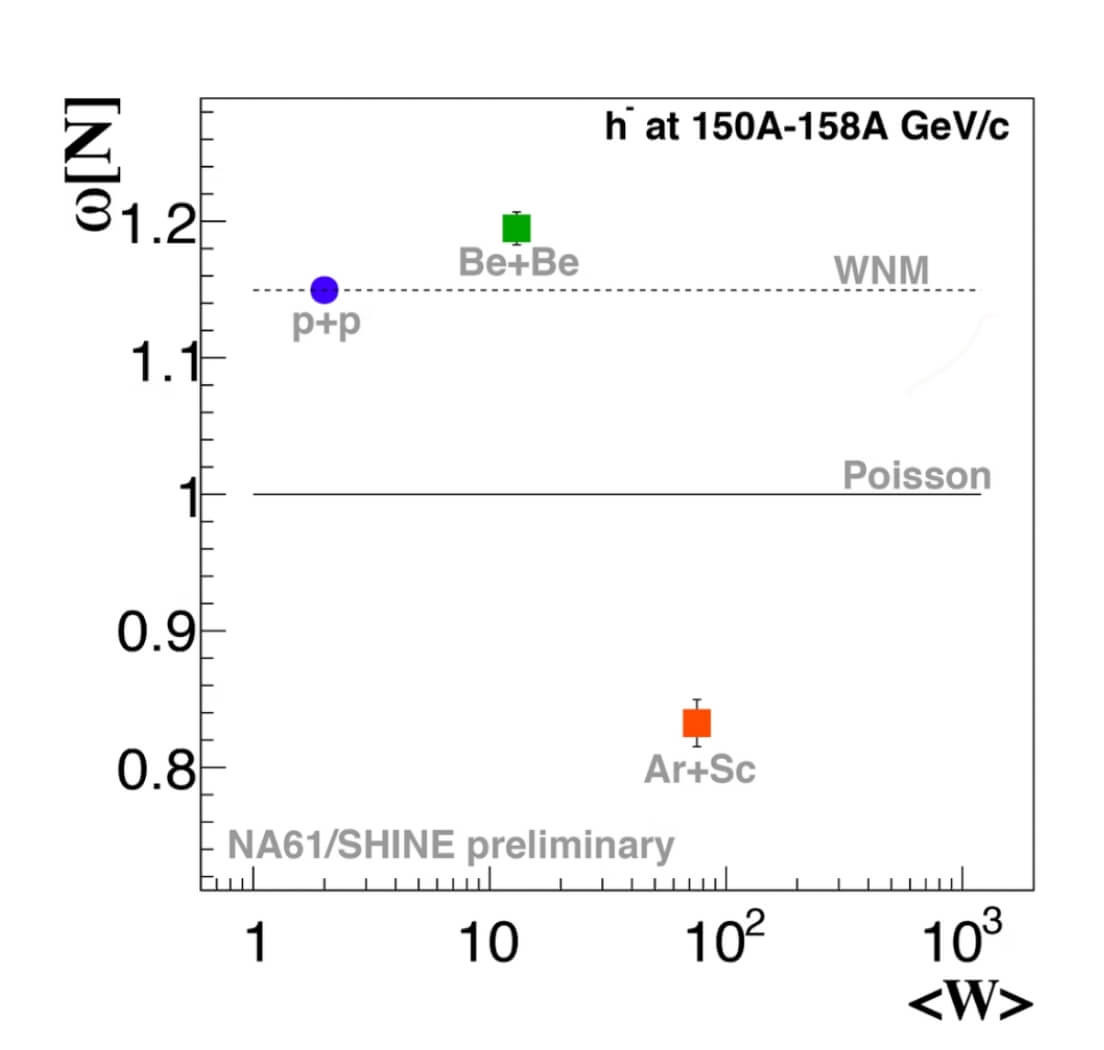}
 \includegraphics[width=0.45\linewidth,clip=true]{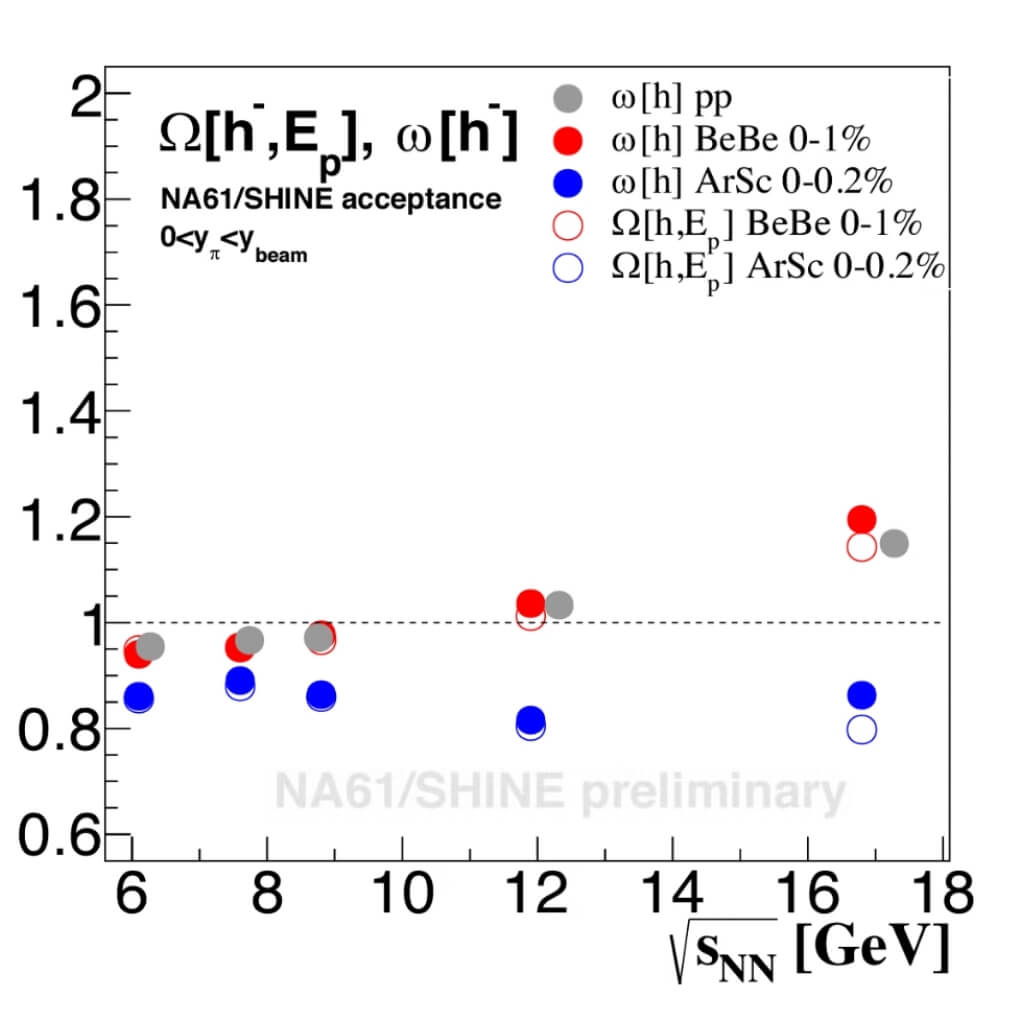}
\caption{
 Measurements of the scaled variance $\omega$ of the multiplicity distribution of negatively charged hadrons in inelastic p+p interactions and central Be+Be and Ar+Sc collisions~\cite{AndreySeryakovfortheNA61/SHINE:2017ygz}: system size dependence at 150\AGeVc (\textit{left})  and  collision energy dependence (\textit{right}). 
 }
\label{fig:cs_oof_results_o}
\end{figure}

\vspace{0.2cm}
\textbf{Unexpected result of measurements.}
Measurements of the system size dependence of hadron production properties at different collision energies were carried out by \NASixtyOne, for detail see Sec.~\ref{sec:cs_ood}. 
Figures~\ref{fig:cs_oof_results} and~\ref{fig:cs_oof_results_o} show the unexpected result~\cite{Podlaski:2019,AndreySeryakovfortheNA61/SHINE:2017ygz}.
The \kp/\pip ratio in Fig.~\ref{fig:cs_oof_results} and the scaled variance of the multiplicity distribution at 150\AGeVc in Fig.~\ref{fig:cs_oof_results_o} are similar in inelastic p+p interactions and in central Be+Be collisions, whereas they are very different in central Ar+Sc collisions which are close to central Pb+Pb collisions. Both reference models, WNM and SM, qualitatively disagree with the data. 
The WNM seems to work in the collisions of light nuclei (up to Be+Be) and becomes qualitatively wrong for heavy nuclei (like Pb+Pb). On the contrary, the SM is approximately valid for collisions of heavy nuclei.  However, its predictions
disagree with the data on p+p to Be+Be collisions.

The rapid change of hadron production properties when moving from Be+Be to Ar+Sc collisions is interpreted and referred to as the onset of fireball. 
From Fig.~\ref{fig:cs_oof_results}~(\textit{right}) follows that the increase of the \kp/\pip ratio depends on the collision energy. On the other hand,
the scaled variance $\omega$ of the multiplicity distribution shows only weak collision energy dependence 
(see Fig.~\ref{fig:cs_oof_results_o}~(\textit{right})).
The physics behind the onset of fireball is under discussion~\cite{Motornenko:2018gdc}.
Whereas one does not observe the formation of fireball in the collisions of light nuclei at the SPS energies, the large size system of strongly interacting matter is probably formed at LHC energies even in p+p high multiplicity events.

\vspace{0.3cm}
\textbf{In summary}, the scans in collision energy and nuclear mass number of colliding nuclei
performed at SPS and RHIC indicate four domains
of hadron production separated by two thresholds: the onset of deconfinement and the
onset of fireball. The sketch presented in Fig.~\ref{fig:future}~(\textit{left}) illustrates this conclusion.


\section{Plans for future measurements}
\label{sec:futures}

\begin{figure}[hbt]

\centering
 \includegraphics[width=0.99\linewidth,clip=true]{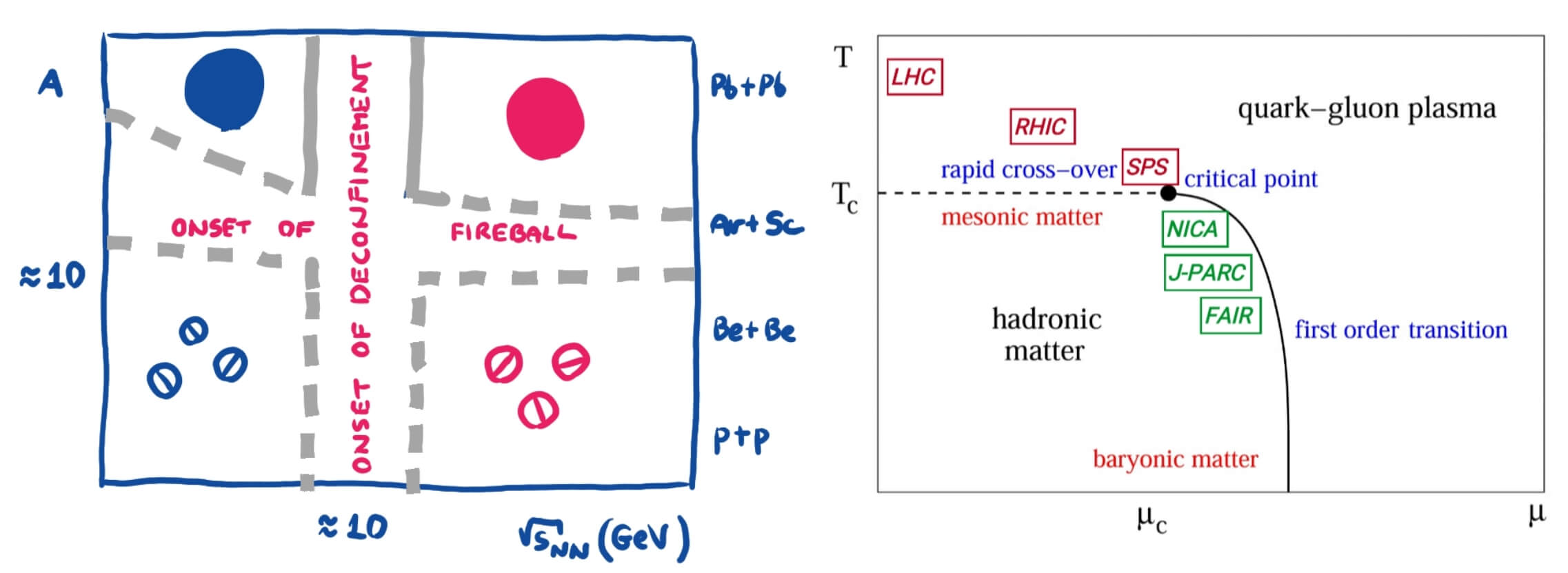}
 \caption{\textit{Left:} 
Two-dimensional scan conducted by \NASixtyOne varying collision energy and nuclear mass number of colliding nuclei indicates four domains of hadron
production separated by two thresholds: the onset of deconfinement and the onset
of fireball. The onset of deconfinement is well established in central Pb+Pb~(Au+Au)
collisions, its presence in collisions of low mass nuclei, in particular, inelastic p+p
interactions is questionable.
\textit{Right:} 
Regions in the phase diagram of strongly interacting matter studied by present (red) and future (green) heavy ion facilities.
 }
\label{fig:future}
\end{figure}

Let us close by discussing possible future measurements which are suggested by this review and which should be considered as priorities: 
\begin{enumerate}[(i)]
\item
A collision energy scan in the onset of deconfinement region to measure open and hidden charm production in Pb+Pb collisions and establish the impact of the onset on the heavy quark sector. 
This requires high statistics data collected with detectors optimized for open and hidden charm measurements. Detailed physics arguments and possible experimental set-ups are presented in Refs.~\cite{Aduszkiewicz:2309890,Agnello:2018evr}.
\item
A detailed study of the onset of fireball and its collision energy dependence in the onset of deconfinement region. The goal is to understand the underlying physics of this phenomenon, for details see Ref.~\cite{Aduszkiewicz:2309890}. This requires a two dimensional scan in the nuclear mass number of the colliding nuclei and in collision energy performed with small steps in nuclear mass number.
\end{enumerate}
Conclusive results from the data recorded by \NASixtyOne and RHIC BES-II are needed to
plan future measurements for the deconfinement-CP search.

Figure~\ref{fig:future}~(\textit{right}) presents a compilation of present and future facilities and their region of coverage in the phase diagram of strongly interacting matter. 
Charm measurements are planned by \NASixtyOne~\cite{Aduszkiewicz:2309890}, NA60+~\cite{Agnello:2018evr} at the CERN SPS, they are considered by MPD~\cite{Kekelidze:2018nyo} at NICA and J-PARC-HI~\cite{Sako:2019hzh} at J-PARC.
A detailed two dimensional scan is considered by \NASixtyOne at the CERN SPS~\cite{Aduszkiewicz:2309890}.

\begin{acknowledgments} 
The work was supported by the the National Science Centre Poland grant 2018\slash 30\slash A\slash ST2\slash 00226 and the German Research Foundation grant GA1480\slash 8-1. The work of M.I.G. is supported by the Program of Fundamental Research of the Department of Physics and Astronomy of National Academy of Sciences of Ukraine.
\end{acknowledgments}

\bibliographystyle{ieeetr}
\bibliography{main}
\end{document}